\documentclass{aa}

\usepackage{graphicx}
\usepackage{txfonts}
\usepackage{newtxtext,newtxmath}

\usepackage[T1]{fontenc}

\DeclareRobustCommand{\VAN}[3]{#2}
\let\VANthebibliography\thebibliography
\def\thebibliography{\DeclareRobustCommand{\VAN}[3]{##3}\VANthebibliography}

\usepackage{graphicx}	
\usepackage{amsmath}	
\usepackage{subfig}
\usepackage{hyperref}
\usepackage{color}
\usepackage{xcolor}

\defcitealias{scott_2020_fornaxI}{Paper I}
\defcitealias{Eftekhari_2021_fornaxII}{Paper II}
\defcitealias{Romero-Gomez_2022}{Paper III}
\defcitealias{Romero-Gomez_2023}{Paper IV}
\defcitealias{McDermid2015}{M15}
\defcitealias{Weisz_2014_sfh}{W14}
\defcitealias{Pasquali2019}{P19}

\newcommand{\alfe}{[$\alpha$/Fe] }

\newcommand{\ppxf}{\href{https://www-astro.physics.ox.ac.uk/~mxc/software/}{pPXF} }
\newcommand{\tng}{\href{https://www.tng-project.org/}{IllustrisTNG} }
\newcommand{\atlas}{ATLAS$^{3D}$ }
\newcommand{\solmas}{M$_{\odot}$ }
\newcommand{\stelmas}{M$_{\star}$ }
\newcommand{\sml}{$\sim$}
\newcommand{\sami}{SAMI-Fornax }

\newcommand{\superscript}[1][]{$10^{#1}$}

\newcommand{\tinf}{t$_{inf}$ }
\newcommand{\tq}{t$_{90}$ }
\newcommand{\qprb}{P$_{Q}$ }

\begin{document}
\title{Are early-type galaxies quenched by present-day environment? A study of dwarfs in the Fornax Cluster.}

\author{
Romero-Gómez, J.,\inst{1,2}\thanks{E-mail: jorgerg658@gmail.com (IAC)}
Reynier F. Peletier\inst{3}
J. A. L. Aguerri,\inst{1,2}
\and R. Smith\inst{4}}

\institute{Instituto de Astrofísica de Canarias, Calle Vía Láctea S/N, La Laguna, Tenerife, Spain
         \and
          Universidad de La Laguna Avda. Astrofísico Fco. Sánchez, E-38205 La Laguna, Tenerife, Spain
          \and
          Kapteyn Institute, University of Groningen, Landleven 12, 9747, AD, Groningen, The Netherlands
          \and
          Universidad Técnica Federico Santa María, 3939 Vicuña Mackenna, San Joaquín, Santiago 8940897, Chile
             }

\date{Received September 15, 1996; accepted March 16, 1997}



\abstract
{Galaxies undergo numerous transformative processes throughout their lifetimes, that ultimately lead to the expulsion of the gas and the cessation of the star-forming activity. This phenomenon is commonly known as quenching, and in this study, we delve into the possibility of this process being caused by the environmental processes associated with the surrounding cluster. 
For this we use the results of \citet{Romero-Gomez_2023}, who analyzed dwarf galaxies in the \sami survey together with massive galaxies from the \atlas survey, for computing the quenching time of each galaxy and compare it with the infall time into the cluster. Using \tq as an approximation for the quenching time and deriving the infall time from phase-space models, we determine the probability of the quenching being produced by the local environment of galaxies.
Our results reveal a relation between galaxy mass and quenching probability. Massive galaxies, down to \stelmas\sml\superscript[10] \solmas, exhibit a low, almost zero probability of quenching, suggesting their independence from environmental effects. As we move into the mass regime of dwarf galaxies, the probability increases with decreasing mass, highlighting their sensitivity to environmental quenching. 
For dwarfs, 36$\pm$9\% of our observational data are consistent with this hypothesis, challenging the idea that the present-day cluster, Fornax, is the primary driver of quenching in the low mass galaxies of our sample with stellar mass from \superscript[7] - \superscript[9] \solmas. 
To further investigate the importance of environmental processes, we compare these results with cosmological simulations, selecting galaxies under similar conditions to our observational sample. Remarkably, the simulated sample shows lower quenching probabilities as we move down in mass, and barely 5$\pm$1\% of galaxies meet the quenching criteria. This discrepancy between observations and simulations underlines that modelling quenching is still in its infancy.
In general, the number of observed galaxies quenched by their environment is lower than expected, which suggests that pre-processing plays a larger role in galaxy evolution. Ultimately, our results highlight the need for higher-quality simulations and refinement of galaxy formation and evolution models.}

\authorrunning{Romero-Gómez, J., et al.}
\titlerunning{Are early-type galaxies quenched by present-day environment?}
\maketitle



\section{Introduction}\label{introduction}
At low redshift, galaxies in the Universe exhibit diverse star formation activity, leading to their classification into two main populations. The first group consists of "star-forming" galaxies characterized by ongoing star formation, young stellar populations, and blue colours, signifying their evolving nature. In contrast, the second category comprises "quenched" galaxies that have ceased to form young stars, displaying older stellar populations and redder colours, this dichotomy is reflected in the so-called colour magnitude diagram \citep[CMD, ][]{Baldry2004_cmd}. The latter population typically displays spheroidal or elliptical morphologies and is frequently found in dense environments like groups or clusters \citep{Binggeli1988, Strateva2001, Baldry2004}. This indicates that the environment can play an important role in galaxy evolution \citep{Boselli2014_enrionment, Boselli2014_}, which can be particularly strong for less-massive galaxies like dwarfs \citep{Binggeli1988}. For this reason, it is usual to characterize an environment with the percentage of quenched galaxies it has, a percentage that depends on the galaxy mass and its environment, since both properties can play a role \citep{Peng2010b, Thomas2010, Vulcani2012, Romero-Gomez_2023}.

For less-massive galaxies, \stelmas < \superscript[10]\solmas, in high-density environments, there are physical mechanisms that can stop the star formation by removing the gas \citep{Lisker2009, Cortese2021}, like ram-pressure stripping \citep{Gunn1972-rampressure, Quilis2000} or strangulation \citet{Larson1980}. Other environment-related processes that can play a weaker role can be harassment \citep{Moore1998harasment, Aguerri-Garcia2009} or even strong interactions with other galaxies that can transform massive galaxies into dwarfs \citep{Moore1999, Mastropietro2005}. In general, for satellite dwarf galaxies falling into massive halos like those of groups and clusters, most of their gas is removed at the first infall before the pericenter of their orbit \citep{Boselli2022}.
On the other hand, galaxies more massive than \superscript[10]\solmas can also suffer strong tidal interactions and mergers that dramatically change the morphology, although in terms of stellar populations, they usually evolve according to their own internal properties without caring too much about the environment \citep{Zheng2019, Davies2019}.

How relevant is the role of the environment? The extent of the environmental influence depends on several factors, the most important are the mass and the dynamical state of the cluster. Fornax is a compact and relatively relaxed cluster, with a lower mass than others, where ram pressure is expected to have a relatively weak effect, in contrast to more massive clusters like Coma or Virgo with a more turbulent nature. In particular, the dynamical state of Fornax suggests that tidal forces may play a more significant role in stripping gas from galaxies, as outlined by \citet{Serra2023}. Conversely, massive clusters such as Virgo, which are characterised by a complex substructure that indicates an unstable and young dynamical state, experience hydrodynamic interactions and galaxy harassment, contributing to the formation of massive early-type galaxies \citep{Boselli2014_enrionment}. Remarkably, for less massive galaxies in Virgo, the increased density of the intracluster medium facilitates the removal of the gaseous component of late-type galaxies through ram pressure stripping events.

As far as observations are concerned: to study galaxy evolution we rely on star formation histories (SFHs) which tell us how much mass (or other related quantities like integrated light) has formed throughout time. Using the SFHs we can infer the impact and dependency on the different properties of a galaxy. In particular, we can investigate when a given galaxy formed its last stars, or, as we say, was quenched. 
Then, it comes into play how the quenching is defined and the properties used to define it, which could depend on the redshift, wavelength or even star-formation-rate tracers used in observations \citep{Speagle2014}. 
Since observational samples could also be affected by selection effects, spectral resolution and integrations \citep{Salim2007, Pintos2019}, the picture that we get from observations is not complete, and therefore comparison with theoretical data could help to solve some uncertainties and better understand the complex process of galaxy evolution.

The aim of this paper is to infer, as a function of the galaxy's stellar mass, the fraction of quenched galaxies from observational data. This will be obtained by computing the time at which 90\% of the stellar population of the galaxies are in place. These observational results will be compared with those resulting from state-of-the-art cosmological simulations of galaxies.

This will be paper V of this series of papers based on the analysis of spectroscopic data of dwarf galaxies, and it is organized as follows: Section \ref{observs-reduct} summarizes the sample selection and spectroscopic observations, section \ref{Data_analysis} we explain the methodology used to analyse the data and obtain the quenching times and probabilities that we present in section \ref{results}. Then we discuss the implication of our results in section \ref{discussion}, and finally, in section \ref{conclusions} we summarize our findings and conclusions.

Throughout this paper we adopt a $\Lambda$CMD cosmology with $\Omega_{m} = 0.3$, $\Omega_{\Lambda} = 0.7$ and $H_{0}$ = 70 km s$^{-1}$ Mpc$^{-1}$.
\section{Data}\label{observs-reduct}
Below we give a brief description of the samples used for this work: The SAMI-Fornax Dwarfs Survey, the \atlas project and the Fornax3D project. This constitutes a sample of 90 galaxies spanning a stellar mass range \superscript[7-12]{}\solmas. More details of the dwarf's sample are in \citet{Romero-Gomez_2022}, hereafter: \citetalias{Romero-Gomez_2022}, while full details of the \atlas and Fornax3D sample can be found in \citet{Cappellari2011-atlas3d} and \citet{Sarzi2018}, respectively.

\subsection{The SAMI-Fornax Survey}\label{obs_seleciton_sec_2_1}
The Fornax Cluster, at $\alpha$(J2000) $= 3^h 38^m 30^s$; $\delta$(J2000) = -35$^\circ$27'18", and a mean recessional velocity of 1454 km/s, it is the second closest galaxy cluster to us, only at a distance of 20 Mpc \citep{Blakeslee2009}. Within this distance, it is the second most massive cluster, with a virial mass of 7x\superscript[13]\solmas, and it is also a relatively small cluster with a virial radius of only 0.7 Mpc \citep{Drinkwater2001}. Fornax is a dynamically evolved cluster \citep{Churazov2008}, with a velocity dispersion of 318 km/s \citep{Maddox2019}. The cluster contains about 1000 known galaxies \citep{Venhola2021}, among which we found a significant population of dwarf galaxies.

For the spectroscopic observations of selected objects, the Multi-Object Integral-Field Spectrograph (SAMI) at the Sydney-Australian Astronomical Observatory (AAO) using the 3.9 m Anglo-Australian Telescope (AAT) was used. The primary targets were 62 early-type, dE, or dS0 galaxies, and the kinematic analysis was performed on 38 of them \citep{scott_2020_fornaxI}. After the spectroscopic inspection, 39 dEs were selected. They are located within the virial radius of the Fornax cluster, except for one associated with the Fornax A group. Some dwarf galaxies showed strong emission lines in their spectra, and these were excluded from the main analysis, leaving a final sample of 31 galaxies. This sample was also analyzed in \citetalias{Romero-Gomez_2022} and in \citet{Romero-Gomez_2023}, hereafter: \citetalias{Romero-Gomez_2023}. This sample of galaxies forms the low-mass tail of the sample, spanning a stellar mass range from \superscript[7.22]{} to \superscript[9.3]{}\solmas.

\subsection{The \atlas project}
To compare with massive galaxies, we include two samples, the first one is located inside the Virgo cluster, and the second is in Fornax. The Virgo Cluster is the closest cluster to us, at a distance of only 16.5 Mpc or a recessional velocity of \sml 1150 km/s \citep{Mei2007}. The cluster has a complex structure with centres around several massive galaxies such as M86, M49 and M87, the former is the most massive \citep{Schindler1999}. As M87 is also the source of the peak emission in X-ray observations \citep{Boringuer_VirgoXray1994}, it is usually considered to be the main centre. Virgo is also a fairly massive and large cluster, with a virial mass of 7x\superscript[14]\solmas \citep{Karachentsev2014} and a virial radius of 1.5 Mpc \citep{Kim2014}. Compared to Fornax, Virgo is a dynamically young cluster \citep{Aguerri2005}, with a velocity dispersion of 753 km/s \citep{Boselli2006}. Its massive galaxies are well-studied systems, and as we know from the literature, the giants are quenched early on, mostly independently of the cluster environment \citep{Heavens2004, Panter2007, Carnall2019,McDermid2015}.

To obtain information on the massive galaxies of the Virgo cluster,] we re-analyzed data from the \hyperlink{https://www-astro.physics.ox.ac.uk/atlas3d/}{ATLAS$^{3D}$ Project} \citep{Cappellari2011-atlas3d}, similar to our previous work. The sample consists of 260 early-type galaxies with a mass between $10^{10}$ and $10^{12} M_{\odot}$, some of which belong to the Virgo cluster, and some are in the field. We divided the sample into clusters and non-clusters based on the local density parameter \citet{Cappellari2011-density}. For this work, we will only use those galaxies that fulfil the cluster membership. We also excluded galaxies with low signal-to-noise ratio, emission lines, or other issues according to \citet{McDermid2015}, leaving a final sample of 50 galaxies. The \atlas galaxies used here were analyzed in \citetalias{Romero-Gomez_2022} and \citetalias{Romero-Gomez_2023}. This sample can be considered a representative sample of massive early-type galaxies.

\subsection{Fornax3D project}
From the literature, we know that massive galaxies barely feel the actions of the environment. Thus, one could assume that massive galaxies from Virgo or Fornax clusters should behave similarly despite the differences between both clusters. Nonetheless, to reassure this, we also add a sample of massive galaxies from the Fornax cluster to compare with the galaxies from Virgo and, on the other hand, with the dwarf galaxies in Fornax.

Fornax3D \citep{Sarzi2018} is a survey that focuses on most of the massive galaxies within the virial radius of the Fornax cluster. Using the Multi Unit Spectroscopic Explorer (MUSE) instrument on the Very Large Telescope (VLT), the survey successfully collected spectroscopic data for a comprehensive sample of 30 galaxies with a limited magnitude of M$_V$ $\leq$ -17 mag. For the purposes of this paper, we use position and velocity data for each galaxy from \citet{Sarzi2018}, along with stellar mass information derived from \citet{Iodice2019_fornax3d} and \citet{Liu2019}.
We also incorporate star formation histories from a subset of the galaxies studied in \citet{Fahrion2021}. They studied 25 objects, from which we have had to remove a couple of galaxies from Virgo, a few dwarfs that were already in our data, one galaxy that is outside of the phase-space range of the models, and other low-mass dwarfs. This leaves us with a sample of ten galaxies that are also part of the Fornax3D project. The selected galaxies are intermediate mass galaxies that have stellar masses ranging from \superscript[9.5] to \superscript[11] solar masses. They are not as massive as the ones in Virgo, but more massive than our dwarfs.  \citet{Fahrion2021} examined galaxy spectra at different apertures; however, for consistency with our treatment of the other datasets, we focus only on integrated spectra at one effective radius. It's worth noting that the methods used to extract star formation histories from the spectra are very similar to our own.

\section{The quenching time of the galaxies}\label{Data_analysis}
To obtain the properties needed in this paper we have to work with the information extracted from the star formation history (SFH) of a galaxy. For a detailed description of the process to obtain the SFHs see \citetalias{Romero-Gomez_2023}, here we just give a summary.

For all the samples, the dEs from Fornax, the galaxies from Fornax3D and the giants from Virgo, the stellar population properties are obtained using full-spectral-fitting techniques, from which we have derived the SFHs. The data in all cases come from integral field spectroscopy instruments, but for the purposes of this paper, we have collapsed all available data into a single spectrum per galaxy. In the case of the Fornax dwarfs, \citet{Eftekhari_2021_fornaxII} analysed the radial profiles and concluded that the velocity dispersion is constant as a function of radius. In this sense, it is safe to assume that the stellar populations do not change much with radius, as dwarfs typically have relatively uniform trends \citep{Koleva2011, denBrok2011, Rys2015, Bidaran2023}. There are some exceptions that can be observed in local clusters. For instance, in some cases, dEs can be identified by the existence of blue nuclei, as noted by \citet{Lisker2006_gradients}. 

For massive galaxies, other studies have also shown that they are characterized by mostly flat age and [Mg/Fe] ratio profiles \citep{Parikh2024}. Even so, in the case of the Virgo galaxies, we tested in \citetalias{Romero-Gomez_2022} the resulting stellar population properties of the integrated spectra at different radii. After careful consideration, we decided to include the entire integrated spectrum of each galaxy in our study, since there was no significant difference between the observed results.

In previous studies of the SAMI galaxies, results were obtained using a spectral range between 4700 and 5400 \r{A}, which is quite similar to that of the \atlas galaxies, between 4800 and 5300 \r{A}. However, for our analysis of the dwarfs we decided to use the red wavelengths of the spectra as well, adding the range between 6300 and 6800 \r{A}. As we shown in \citetalias{Romero-Gomez_2023}, this extended wavelength range helped us to improve the results of the kinematics that are necessary to obtain the stellar populations, but it does not have any impact on the populations.

All the fitting and analysis of the galaxy spectra is made with the code \ppxf \citep{Capellari2004, Capellari2017}. The spectrum of each galaxy is fitted with a combination of single stellar population (SSP) models from the \href{http://miles.iac.es/}{MILES} library \citep{Falcon-Barroso2011, Vazdekis2010, Sanchez-Blazquez2006}. The analysis of the Fornax3D galaxies in \citet{Fahrion2021} uses the same code and SSP models. The age of these models can be seen as the time at which the different stellar populations formed, thus the SFH is composed of the weights of the SSP at different times. The quality of our spectra in both dwarfs and giants ensures that the \ppxf fit is good. The time resolution of our fit is about 1-2 Gyr, the maximum allowed by the data. This means that we do not have enough time resolution to distinguish individual bursts, and for this reason, we focus on the overall shape of the reconstructed SFHs, rather than on small features that may appear (see \citetalias{Romero-Gomez_2023} for more details).

\subsection{Quenching and infall times}\label{explain_regul}
The star formation history is often presented as a function of galaxy age, showing the mass of stars formed at a given point in the galaxy's evolution. For our research, we are interested in the amount of mass formed up to a given point in time, and for this, we have used the cumulative SFH. This function can be easily constructed as the cumulative sum of the resulting weights of the fit. This allows us to calculate, for any given time, the percentage of a galaxy's stellar mass that has been formed relative to the total stellar mass at the present time. In particular, we are interested in knowing when a given galaxy reached a fraction equal to 90\% of its present mass. This parameter, the \tq, can be considered as a proxy of the quenching time of the galaxy, when the star formation has entirely ceased \citep[see][]{Weisz2014, Ferre2018, Collins2022}. For example, since we are working in lookback time, the smaller the \tq is, the longer the galaxy has been forming stars. 
One has to realize that the \tq parameter is dependent on the time resolution of the SFHs.

\begin{figure}[!h]
\centering
\includegraphics[scale=0.845]{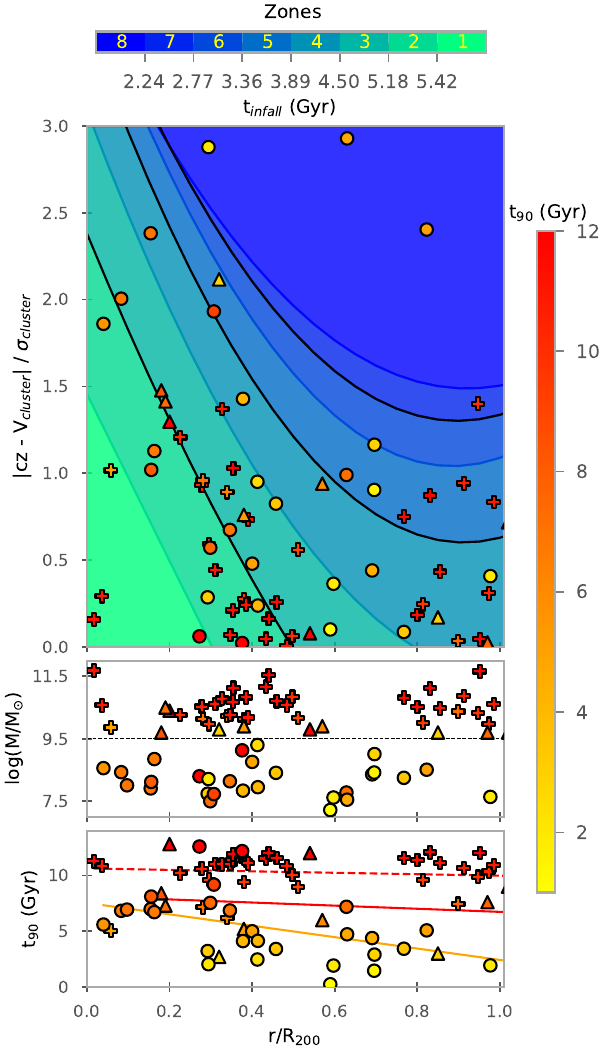}
\caption{Plots showing the distribution of the galaxies in the phase-space, and some of their properties. The top panel shows the phase space for the \sami, \atlas and Fornax3D samples. This phase space is divided into eight different zones, according to \citet{Pasquali2019}, that represent the infall time of the galaxies in those zones. For a better contrast, we marked with black lines zones 2, 4 and 6. From green to blue the infall times in Gyr change according to the top color bar. The middle panel shows the stellar mass of the galaxies as a function of the projected distances, with a horizontal black dashed line that separates the Fornax's dwarfs from the rest of the galaxies. In the bottom panel, we present the \tq of the galaxies as a function of the projected distances, with a dashed red line fitting the trend for giants of Virgo, a red line for the Fornax3D galaxies and an orange line fitting the trend for dwarfs. These relations of \tq with the environment were already highlighted in \citetalias{Romero-Gomez_2023}. In all panels, we show the \sami dwarfs with circles, while the massive galaxies for Virgo and Fornax are marked with resp. crosses and triangles. All symbols are coloured with \tq, which value is indicated by the vertical colour bar.}
\label{fig_1}
\end{figure}

The second time scale that we define is the infall time, \tinf, a property that does not depend on the internal properties of the galaxy, only on the environment. This time is defined as the moment when a galaxy that is falling into the cluster crosses the virial radius for the first time. Since we cannot go back in time with the observed galaxies, we required simulations to help us with this. In \citet{Pasquali2019}, the YZiCS simulations \citep{Choi_Yi2017} are used to study the cluster infall time of galaxies in halos from 5.3x\superscript[13] \solmas to 9.2x\superscript[14]\solmas. To do this they used the phase space, which is a representation of the position of the galaxies within their parent halo. This is achieved by plotting the peculiar velocities, normalized by their host (cluster) velocity dispersion, as a function of their projected distance, normalized by the host virial radius R$_{200}$. Thus, the phase space gives us some constraints on the infall time of the galaxies. However, in phase space, there can be potential biases due to projection problems. Therefore, given a galaxy's position in the diagram, a range of infall times can be derived, each of which is associated with a probability. In this sense, is reasonable to think that the closer to the centre a galaxy is, the longer it should have been within the host. Therefore, \citet{Pasquali2019} divided the phase space into 8 zones using quadratic curves, and each of these zones is represented by the mean infall time of all the galaxies within. The different zones can be seen in the phase space of Figure \ref{fig_1}. Using these models and the position of our galaxies in phase space, we obtain the infall time of each galaxy in our sample as the mean of the corresponding region of phase space in Fig. \ref{fig_1}. This means that this time is an approximation in the sense that, given the position, it is the most likely time according to the models. Even though the models use galaxies slightly more massive than our dwarfs, the \tinf from the models are not very different for the galaxies in our mass range. Thus, since we do not actually know the infall time of a given galaxy, we assume it is the mean \tinf of its zone given by \citet{Pasquali2019}.
\subsection{Environmental Quenching Probability}\label{probabilty}
How can we know if a galaxy has been quenched by its environment? Well, we cannot be 100\% sure that a host halo is responsible for quenching its satellites, however, if the quenching time of a galaxy is shorter than its infall time, then there is a higher probability that the quenching is related to the environment. With this premise, and taking into account that our dwarf galaxies are quenched by selection, we defined the probability that they were quenched in the cluster, P$_{Q}$, the Environmental Quenching Probability. To compute this value we look whether:
\begin{equation}\label{def_qprb}
    t_{90} - t_{inf} < 0
\end{equation}

This essentially means that the infall time is longer than the quenching time, both measured in look-back time. Then, using the errors computed for the \tq in \citetalias{Romero-Gomez_2023} and for the \tinf in \citet{Pasquali2019}, we did Monte-Carlo simulations to check how many times the condition of equation \ref{def_qprb} was fulfilled. Thus, we interpreted the percentage as the probability of a galaxy being quenched by its environment. 
Under these conditions, the limit to consider whether a galaxy has been influenced by its environment is when \tq - \tinf = 0.  It is important to note that we are assuming that the quenching time we are measuring is primarily influenced by the environment. We have not considered the possibility of mass quenching, as well as merging galaxies, which are more appropriate for massive galaxies and that could also may stop the star formation activity \citep{Peng2010b}. In section \ref{discussion}, we will delve into this further and discuss how it may affect our results.


\section{Results}\label{results}
Figure \ref{fig_1} displays the phase-space distribution of all the galaxies within their respective clusters, note that in the case of the \atlas sample we used only those galaxies in a cluster environment \citepalias[for more details see][]{Romero-Gomez_2022}. It reveals intriguing insights into their respective distributions. The phase space is divided into eight distinct zones, according to \citet{Pasquali2019}, with zone 1 representing the region closest to the centre, while the zones progress outward in increasing distance. When considering the overall trend, it becomes evident that both dwarfs and massive galaxies exhibit distinct patterns within the phase space. The respective percentages of galaxies in each zone are given in Table \ref{table_1_objects_info}. We see that the dEs galaxies in Fornax demonstrate a relatively uniform distribution, with a significant proportion of objects occupying zones 2 to 5 in Fig. \ref{fig_1}, between 18 and 25\% in each zone. For the small sample of Fornax3D galaxies, we see that all of them are between zones 2 and 5 with the majority, 40\%, in zone 3.
On the other hand, the giant galaxies in Virgo exhibit a more concentrated distribution, with 75\% of objects found between Zones 2 and 4. Additionally, there is a notable difference in the outermost cluster regions, with the dwarf sample showing a notable presence in Zone 8, which contains 11\% of objects are there. Of the giants, only 12\% of objects are in zone 5, and only one giant galaxy is in zone 7. The virialized region, zone 1, is less populated. Both clusters show similar numbers, 4 and 9\% for dwarfs in Fornax and giants in Virgo, respectively. This is expected to some extent, as can be seen in Fig. 5 of \citet{Pasquali2019}, the highest density of objects in the models is in zones 2 and 3. This is produced because while galaxies are orbiting the cluster, the passage through the central parts is fast, making it less probable that we observed them in those regions.

Also in Fig.\ref{fig_1}, we show the distribution of stellar masses as a function of the projected distances. This does not show any relation between these two properties, but it clearly shows that dwarfs and massive galaxies are well separated in stellar mass. All the dEs are below \superscript[9.5] \solmas, and all the giants are above. For the \tq of the galaxies, as pointed out in the literature \citep{Sandage1986, Gavazzi1996, Romero-Gomez_2023}, the values are strongly correlated with stellar mass, in the sense that massive galaxies formed in 2-3 Gyr after the Big Bang while the dwarfs have been forming slower over time. As for the relation between \tq and the distance, it only exists for dwarf galaxies, as already pointed out in the literature \citep{Michielsen2008}. Showing that those dEs closer to the centre of the cluster were quenched faster.

\subsection{Galaxy quenching}
As described in the previous section, the difference between \tq and \tinf gives us the definition of the Environmental Quenching Probability. Considering a galaxy to be quenched by the environment if \tq is equal or lower than \tinf, after analyzing the distribution of \qprb as a function of \tq - \tinf we found that the limit condition of quenching, \tq - \tinf = 0,  corresponds to \qprb = 0.497 $\pm$ 0.093. We find that 3$\pm$3\% of the Virgo giants, 36$\pm$9\% of dwarfs and 20$\pm$13\% of the intermediate massive galaxies in Fornax have probabilities higher than this limit, so in total, only 18$\pm$5\% of the whole sample is compatible with being quenched by its present-day environment. Narrowing our focus to the Fornax cluster, dwarfs play a dominant role in the statistics, with about 31$\pm$7\% of Fornax galaxies showing that they are compatible with quenching. Notably, there are differences between the samples of massive galaxies with those within the Fornax cluster showing a higher propensity for quenching influenced by their surroundings. 
\begin{table*}
\caption{Table with the percentages of galaxies on each zone of the phase-space. The simulated galaxies have been divided into two samples according to their stellar mass in order to easily compare with the dwarf and giant samples.} 
\centering    
\begin{tabular}{ccccccccc}     
\hline
Zones & 1 & 2 & 3 & 4 & 5 & 6 & 7 & 8 \\

\noalign{\smallskip}
\hline\noalign{\smallskip}

\noalign{\smallskip}
\sami dwarfs & 4\% & 25\% & 25\% & 18\% & 18\% & 0\% & 0\% & 11\% \\

\noalign{\smallskip}
\atlas & 9\% & 30\% & 24\% & 21\% & 12\% & 0\% & 3\% & 0\% \\

\noalign{\smallskip}
Fornax3D & 0\% & 10\% & 40\% & 30\% & 20\% & 0\% & 0\% & 0\% \\

\noalign{\smallskip}
TNG50 (M $\leq 10^{9.5}$) & 22\% & 28\% & 22\% & 18\% & 5\% & 2\% & 1\% & 1\% \\

\noalign{\smallskip}
TNG50 (M $> 10^{9.5}$) & 28\% & 31\% & 24\% & 9\% & 6\% & 2\% & 0\% & 0\% \\

\hline\noalign{\smallskip}

\end{tabular}
             
\label{table_1_objects_info}
\end{table*}
\begin{figure}
\centering
\includegraphics[scale=0.7]{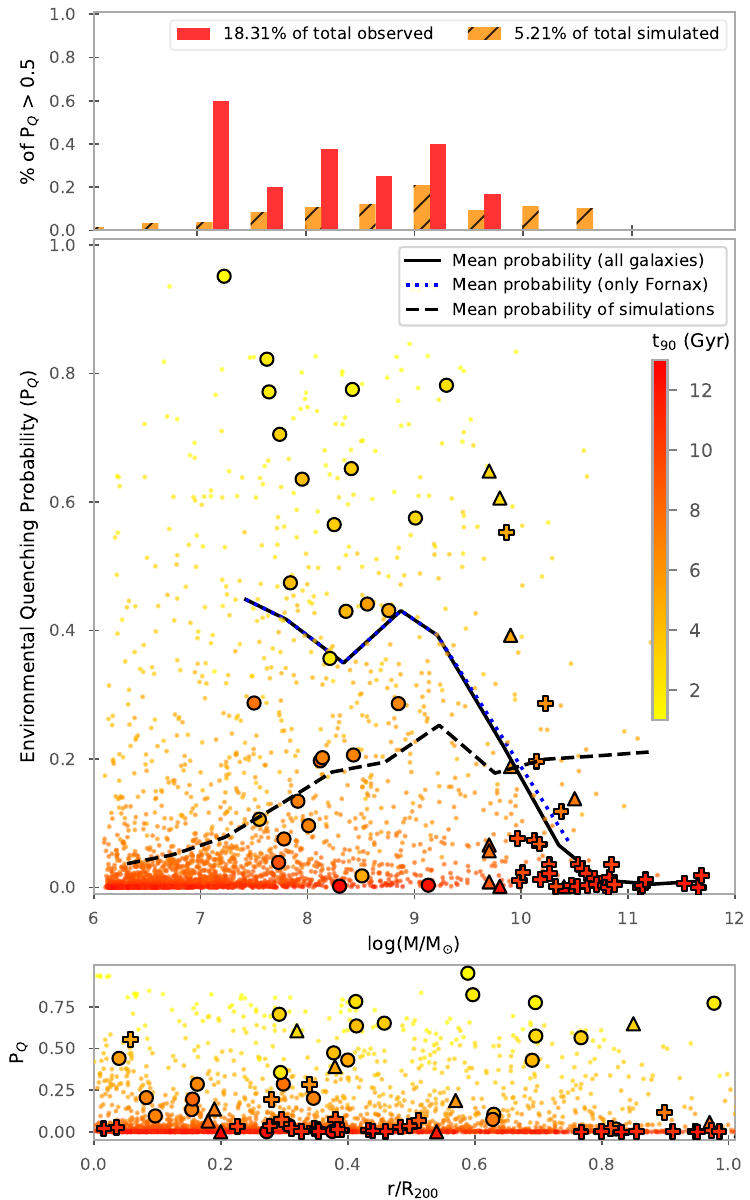}
\caption{Plots with the relation between the Environmental Quenching Probability and the stellar mass. In the top panel, we show a histogram with the percentage of quenched galaxies in different mass bins, separated by observed galaxies in red and simulated galaxies in orange. In the central panel, we present the individual values of \qprb as a function of the stellar mass, with symbols coloured by the \tq. In the bottom panel, we show the distribution of \qprb as a function of the projected distance. For the observational sample, the symbols are the same as in Fig. \ref{fig_1}, and the background points are the simulated galaxies from the Illustris Project. For all the observed galaxies we include a black line that represents the mean \qprb in different mass bins, and with a blue dotted line we represent the same fit but only for the galaxies of the Fornax cluster. For the simulated galaxies the mean \qprb is represented by the dashed black line.}
\label{fig_3}
\end{figure}

Given that \tq is related to the stellar mass, it seems logical that \qprb is related too. For that reason in Fig. \ref{fig_3} we present a compelling view of the relationship between the quenching probability and the stellar mass. The results reveal a distinct trend in the quenching probability as a function of stellar mass. For high mass, the probability remains consistently low, nearly zero, until reaching stellar masses around \superscript[10] \solmas. These galaxies have been quenched by internal processes. At this point, the probability starts to rise up to \qprb = 0.4, which is more or less maintained until \superscript[8] \solmas. Upon closer examination of the Fornax3D galaxies, most of them have \qprb < 0.5, which means that their environment most likely did not play a major role in the quenching. 
For lower stellar masses we reach the sample of dwarf galaxies in Fornax, note, that some galaxies with almost zero probability, while for some others we see that the quenching probability increases almost exponentially and reaches almost \qprb = 1.

Comparing the galaxies between all the samples, the results indicate that the massive galaxies samples predominantly exhibit low quenching probabilities. Approximately 94\% of the giants fall below \qprb = 0.2. In contrast, the dwarf sample displays a different distribution, with around 68\% of the dwarfs surpassing the 0.2 quenching probability. These findings emphasize that the quenching probability of dwarfs and giants is different, showing that, as expected, low-mass galaxies are much more sensitive to the environment and thus, more likely to be quenched by it. It is also consistent with giants self-quenching, independent of the environment \citep[as can be seen in][]{Peng2010b, Pasquali2019, Smith2019}{}{}.

\subsection{Quenching of Simulated galaxies}
To further investigate this and test the consistency of our findings, we now compare our results with those of simulated galaxies. For the simulated sample we used the results obtained from the \tng project \citep{Marinacci2018, Naiman2018, Nelson2018, Pillepich2018b, Springel2018}, which comprises cosmological magnetohydrodynamical simulations conducted within different comoving volumes. These simulations incorporate various physical processes relevant to galaxy formation and evolution like star formation and evolution, gas heating and cooling, black hole growth through mergers and accretion, and feedback mechanisms such as galactic winds and AGN activity. Detailed information about the physics implemented in the TNG simulations can be found in \citet{Weinberger2017} and \citet{Pillepich2018a}. All TNG runs assume a flat $\Lambda$CDM cosmology with parameters derived from the \citet{Planck2016} results. In our study, we specifically focused on the TNG50 simulation \citep{Nelson2019b, Pillepich2019}, which provides the highest resolution within the suite, a mass particle of 8.5x\superscript[4]\solmas. 

To have a sample of simulated galaxies as similar as possible to our samples of observed galaxies, first, we select only those halos with a virial mass as low as the Fornax cluster, and no more massive than the Virgo cluster, 13.5 < log(M$_{200}$/\solmas) < 15. These halos are slightly more massive than Fornax, but with this selection, we can ensure that the simulated clusters have a good statistical and numerous samples of galaxies. Then, since our galaxies do not show strong signs of active star formation, we impose a condition on the gas-to-star particle ratio, so that we only select galaxies with a maximum of 20\% of gas, so that can be considered as early-type. And finally, for the resolution, since we are interested in the SFHs, we include on the final sample only galaxies with 50 star particles or more, to ensure that we have enough resolution \citep{Alberto2023}. These filters resulted in a sample of 2677 simulated galaxies with 6.1 < log(\stelmas/\solmas) < 11.3.

After the selection, we employ the same methodology as in the observed galaxies. Obtain the SFH and compute the \tq to use it as quenching time, place the galaxies in the projected phase-space to use \citet{Pasquali2019} zones to get the \tinf, and then with those values compute the \qprb for each simulated galaxy. The different percentages of the distribution in the phase-space can be seen in Tab. \ref{table_1_objects_info}, while the relation between \qprb and the stellar mass is represented in Fig. \ref{fig_3}. Like for the giants and dwarfs, the simulated galaxies are also concentrated in zones 2 to 4. However, contrary to the observed sample that is barely present in the virialized region zone 1, 22\% of the simulated galaxies are in this zone. Although there are some simulated galaxies with high \qprb, looking at Fig. \ref{fig_3} we see that the mean trend of the simulations gives a completely different result. The majority of the galaxies below \superscript[10] \solmas are not compatible with being quenched by their current environment, only 5$\pm$1\% are quenched by the present-day environment. These opposite results raise some interesting questions about the evolution of galaxies in clusters. To look deeper into the differences, we divided the simulated sample depending if a galaxy's stellar mass is bigger or smaller than \superscript[9.5]. The simulated galaxies that are more massive, similar to the giant sample, are more concentrated in the central regions and are also more quenched, 11$\pm$3\%. On the other hand, the less massive simulated galaxies that can be compared to the \sami dwarfs are as quenched as the giants from \atlas, 5$\pm$1\%. This behaviour is visible in Fig. \ref{fig_3}, where we fitted a line to the mean values of different mass bins and the relation goes down as we go down in mass.

\section{Discussion}\label{discussion}
The findings presented in our study shed light on the relationship between the probability of local environmental quenching and stellar mass for observed galaxies. 

\subsection{Infall time and the goodness of the sample}
When we incorporate the simulated galaxies into Fig. \ref{fig_3}, the first thing we notice is that the extent of \qprb is quite similar for both observed and simulated galaxies. This indicates that the quenching times reproduced by the simulated physical mechanism are in the same range as the observed galaxies. Looking at the mean \qprb of the simulations, we observe that giant simulated galaxies (\stelmas > \superscript[9.5]\solmas) exhibit slightly higher quenching probabilities in the simulations, which gradually increase until a stellar mass of \sml\superscript[10]\solmas. This aligns with our observational results and suggests a degree of consistency between the two datasets. However, a notable departure from the observed trend emerges when examining less massive galaxies in the simulations. Contrary to the observations, the mean probability for a galaxy to be quenched decreases for lower-mass galaxies. This discrepancy can be attributed to the presence of a significantly larger number of dwarf galaxies with higher quenching times in the simulated sample, i.e. dwarfs that were quenched quickly after the Big Bang. As a result, they are generally somewhat bluer. These high-quenching-time dwarf galaxies exert a downward influence on the mean quenched probability, leading to the observed decline. In the \sami sample we find dEs with similar \tq, but in such small numbers that they do not have a significant impact on the mean trend. 

\begin{figure}
\centering
\includegraphics[scale=0.82]{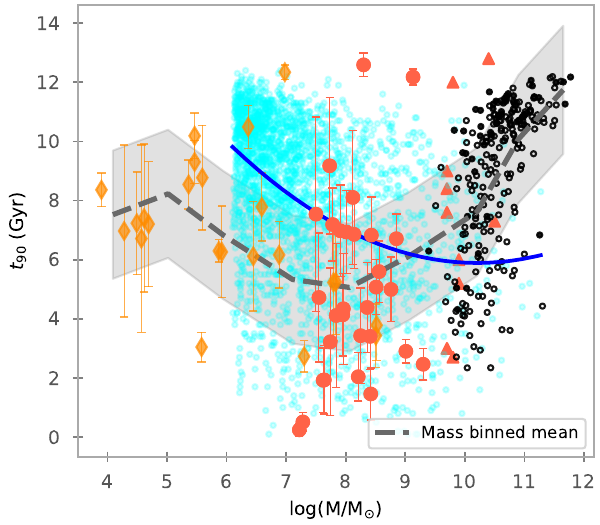}
\caption{Figure showing the relation between \tq and the stellar mass, as presented in \citetalias{Romero-Gomez_2023}. The orange diamonds are dwarf galaxy satellites of the Milky Way or Andromeda, these galaxies are included in the plot just to show how the U-shape continues for less massive galaxies. The red circles are the dwarfs from the \sami sample. The giant galaxies from the \atlas project are represented as empty or filled black dots depending if they are in the field or a cluster, respectively. In addition to the figure in \citetalias{Romero-Gomez_2023}, we have added the Fornax3D galaxies as red triangles. For all the observed galaxies we include the mass binned mean of \tq with a dashed grey line and grey shadow for the standard deviation. The simulated galaxies from \tng are the cyan points, while the blue line is the mass binned mean of these galaxies.}
\label{fig_4}
\end{figure}

To further investigate the results of the simulations, we try to replicate the U-shape of \citetalias{Romero-Gomez_2022} and \citetalias{Romero-Gomez_2023} with the simulated data. This is a special shape related to galaxy evolution that appears when studying properties that depend on internal and external properties. One such property is the abundance ratio \alfe, which can be obtained as part of the different stellar population properties resulting from the full-spectrum fit. The $\alpha$ elements, such as Mg or C, are produced in Type II supernovae, whereas Fe is predominantly produced in Type Ia supernovae. The former marks the end stage of stars with relatively short lives, while the latter originates from progenitors with significantly longer lives. Therefore, $\alpha$-element abundance values can serve as a proxy for the timescales of star formation in a galaxy \citep{Reynier1989PhD, Worthey1992}. When comparing the \alfe values of dwarfs and giants as a function of stellar mass, they form a U-shaped curve. The relation for massive galaxies is mostly linear, indicating that they evolve based on their internal properties. For galaxies with a stellar mass lower than \sml\superscript[8], the relation begins to rise, and the higher values of \alfe correspond to low-mass galaxies that are closer to their host. This demonstrates how the evolution of dwarf galaxies can be influenced not only by their internal properties but also by their environment, which can rapidly suppress the star formation of less massive galaxies. In accordance with this, our previous study \citepalias{Romero-Gomez_2023} demonstrated that another parameter associated with galaxy evolution, the \tq, also exhibited a U-shaped distribution. In general, this distribution shows how dwarf galaxies with a stellar mass of \sml\superscript[8]\solmas are formed more slowly, showing the lowest star formation rates. In Fig. \ref{fig_4} we replicate the U-shape from \citetalias{Romero-Gomez_2023} between the \tq and the stellar mass, and include the simulations in the picture for comparison. The figure also includes dwarf galaxies from the Local Group, in order to show how the relation behaves for lower masses as well as to manifest how the U-shape nicely connects even for galaxies in different environments. 
Although weak and diluted, the simulated galaxies also show a possible U-shape. However, this simulated U-shape has the minimum \tq centred around \superscript[10]\solmas, two orders of magnitude higher than the observations. This suggests that the physical mechanisms acting in the simulations are making intermediate mass dwarfs older than the observations, while the giant galaxies are younger than they should be. Our findings are consistent with the analysis that \citet{Joshi2021} made of the cumulative SFHs of dwarfs in TNG50 across a variety of environments. They described the weak U-shape as a decreasing of \tq with satellite mass and then a flattering around \superscript[9-10]\solmas. This and the behaviour of the SFHs are strongly related to stellar mass, as they found, and it is also dependent on the host mass, the position inside it and the accretion time. They conclude that the phase-space made by \citet{Rhee2017_caustic} can be used to condense all of this, separating the satellite galaxies into ancient, intermediate and recent infallers. 

Similar to \citet{Pasquali2019}, \citet{Rhee2017_caustic} parameterized the phase-space into different zones and for each one gave the probability of being ancient, intermediate or recent infaller, along with the corresponding infall time. Using the probabilities to compute mean infall times for each zone results in similar numbers to those of \citet{Pasquali2019}, but separating them into infall groups allows the infall time to be older. For example, the galaxies closer to the centre in Fig. \ref{fig_1} are given an infall time of 5.42 Gyr, but since they are probably ancient infallers, their infall time could be 8.35 Gyr according to \citet{Rhee2017_caustic}. In \citet{Joshi2021} the accretion time can be as old as 12 Gyr, which is more similar to the ancient infallers time.

\citet{Ding2023} investigated galaxy infall times, with a particular focus on intermediate-massive galaxies within the Fornax cluster. The methodology involved determining the galaxy infall time by establishing a robust correlation with the cold-disk age, a relationship calibrated using the TNG50 cosmological simulation. The study found that the resulting times were statistically aligned with the positions of the galaxies in the phase space. This study has 3 dwarfs in common with our paper: FCC143, FCC 182 and FCC 301. However, in each case, the cold disk fraction is only at most 0.07, so that their results cannot be compared with our results here. In any case, they used the phase space division outlined by \citet{Rhee2017_caustic}, which as said before have infall times larger than those from \citet{Pasquali2019}. With those numbers the \qprb would indeed be also larger, giving us more galaxies compatible with the environmental quenching.

This highlights how the probability of environmental quenching may vary depending on the measurement method used and means that the results presented in this paper are dependent on the models used. However, it is important to note that the accuracy of these models depends on the correlation between the position in phase space and the difference between recent and old infallers. Therefore, the reliability of this correlation within the adopted models is crucial for the robustness of the results.

With all this in mind, the results using the zones and times provided in \citet{Pasquali2019} are to be taken with caution since we assume the infall time of a galaxy is equal to the mean for that zone, and not its real infall time. The results should not change that much if we could use the real \tinf, because if we had enough galaxies they would be closer to the mean. This means the fractions that we measure are probably very noisy due to the relatively low number of objects in our observational sample. Thus, our observational sample may not represent a good statistical sample and we could have a selection bias in the dwarfs regime. In this regard, the observational sample it is just a single line of sight and could be affected by projection effects and cluster asymmetry. Meaning that with more observations the statistics would be compatible with the simulations. To test this hypothesis we tried to replicate the stellar mass distribution of the observational data with the simulations. Using mass bins of 0.25\solmas, we randomly selected simulated galaxies so that the ratio between bins and the total number of galaxies was equal in both simulated and observational samples. Once the distribution was the same in both samples, we replicated Fig. \ref{fig_3} with the randomly selected galaxies. The trend between between \qprb and stellar mass was almost identical, hence, the target selection for the observation does not seem to have any notable impact on the results. 

The problem, instead, could be that the simulations do not produce realistic quenching for a chosen environment \citep[see ][]{Xie2020}. The observed fractions of quenched satellite galaxies are difficult to reproduce using theoretical models of galaxy formation and evolution. In the past, state-of-the-art models tended to overestimate these fractions, especially in the low-mass regime \citep{Weinmann2006, Baldry2006}. The quenching times predicted in these models were only about 1 Gyr, which is significantly shorter than the several Gyrs inferred from the observational data \citep{Wetzel2013}. Comparison of observational data with semi-analytic models leads to a similar underestimate of the quenching time \citep{DeLucia2012}, though is clear that the timescales needed decrease with increasing stellar mass \citep{Hirschmann2014}. This difference could be a fundamental source of disagreement between our results and the observed and simulated galaxies.

\subsection{Quenching time and pre-processed galaxies}

To study the relation between quenching time and pre-processing in the TNG50 simulations, \citet{Joshi2021} defined the quenching time as the moment when the star formation rate of a galaxy is 1 dex below the corresponding star-forming main sequence. Their results showed that using \tq as an approximation of the quenching is reasonably valid, although there could be a delay between the time when galaxies assemble 90\% of their stellar mass and the moment of quenching. This is caused by the little bit of star formation left in the galaxies.
 
This bias arises from the fact that our results are based on the assumption that the quenching is dictated by the environment, therefore a galaxy is quenched after its infall in a cluster or group \citep{Muzzin2008, Fillingham2015_108_mass_therhold}. From the study of massive galaxies, we know that the mass formation inside a galaxy can expel the gas contributing to the quenching of the galaxy. Because of this, we are biased if there has been a process of mass quenching in the galaxy. So, does this mean that the dwarf galaxies are not quenched by the environment?

To answer the question, \citet{Peng2010b} studied galaxies in different environments using data from surveys such as SDSS \citep{York2000} and zCOSMOS \citep{Scoville2007}. They concluded that the quenching of massive galaxies was primarily due to mass-quenching processes rather than environmental processes. However, for satellite galaxies, their results indicated that between 30\% and 70\% of them are quenched when they fall into a larger halo. These percentages are consistent with our findings in this study of Fornax dwarfs, in which 36$\pm$9\% of them are most likely quenched by the environment\citepalias{Romero-Gomez_2022}. According to \citet{Serra2023}, ram-pressure stripping alone is not sufficient to quench less massive galaxies in the Fornax cluster, and tidal forces are also necessary. This finding is consistent with the observations of \citet{Kleiner2023}, who studied the HI content of Fornax galaxies. They detected HI in less than \sml5\% of their sample and found that most of the HI-detected dwarfs were not close to massive galaxies or the cluster centre. They concluded that massive galaxies also play an active role in removing gas from dwarfs. Looking at the position of our dwarf galaxies in Fornax (see Fig.1 in \citetalias{Romero-Gomez_2022}), most of them are close to galaxies with stellar masses greater than \superscript[10] solar masses. In addition, some of our dwarfs have \tq much longer than their \tinf obtained from the phase space. This adds another cause to environmental quenching and introduces the possibility that some dwarfs may have fallen into the cluster as part of a group and arrived already quenched by the effect of that group, a phenomenon known as pre-processing \citep{Bidaran2022}. There is also the possibility that some galaxies, which are located on the edges of the phase space and have long quenching times, may be backsplash galaxies \citep{Sales2007}. These galaxies fell into the cluster a long time ago, but after passing near the core, their orbits create a slingshot effect that throws them to the outskirts of the cluster. During this time, the environment can quench the galaxy. However, it is difficult to distinguish these galaxies from those that are falling into the environment through observations \citep{Pimbblet2011_blacksplash}, and their study is more common in $\Lambda$CDM cosmological simulations of clusters \citep{Haggar2020}.

To investigate the effect of the environment on the quenching of galaxies in simulations, a detailed study of various quenching pathways was conducted by \citet{Donnari2021a} using TNG300. These are the lowest resolution versions of the TNG models, meaning that the stellar mass of galaxies is \superscript[9-12]\solmas.
Among the pathways studied by \citet{Donnari2021a}, pre-processing generates an important fraction of quenched galaxies. For example in massive hosts, M$_{200}$ > \superscript[14.5]\solmas, only half of the low-mass satellites are quenched in that environment. These numbers increase as the galaxy's stellar mass goes down, which suggests that the numbers could be even higher for the TNG50 galaxies of \superscript[6-7]\solmas. In general, around 30\% of the satellites found in clusters experienced quenching before becoming part of their current host at z = 0. This would mean that the quenching probability obtained from the phase-space at z = 0 could not be a good test to know if a galaxy has been quenched by its current environment. \citet{Donnari2021a} also noted that compared to observations, the number of quenched galaxies in TNG is higher, which they argued is because of a combination of different physical mechanisms. In the IllustrisTNG simulations, this problem is known as the over-quenching effect \citep{overquenching_problem}. It means that environmental quenching is not only due to one host, the quenching process could start in a sub-group and finally halt quickly after arriving at their current host. Looking at the definition of the quenching probability, the aforementioned process would also add galaxies with low \qprb in Fig. \ref{fig_3}. In general, they found that the fraction of quenched low-mass satellites is higher in more massive hosts like clusters, and also higher near the host centre, which is expected in the environmental quenching scenario. While galaxies with \stelmas > \superscript[10]\solmas are only quenched by internal processes like AGN feedback, independently of host mass, position or infall time, agreeing well with our results in \citetalias{Romero-Gomez_2023} and Fig. \ref{fig_3} of this paper.

\section{Summary and Conclusions}\label{conclusions}
In this paper, we present a study of the probability that galaxies have been quenched by their present-day environment. For that purpose, we used the \tq obtained from the SFHs as an approximation of the quenching time and derived the infall time of each galaxy from the parametrization of the phase space. As observational datasets, we use a sample of 31 dEs from the \sami and 50 ETGs of the ATLAS$^{3D}$ project. Also, for comparison, we analysed the same properties on a similar sample of simulated galaxies from the Illustris TNG50.

\begin{enumerate}
  \item Our results demonstrate robust agreement with previous findings in the literature. Specifically, more massive galaxies evolve on their own, while less-massive galaxies, such as dwarfs, can experience quenching due not only to their internal properties but also as a result of environmental processes.
  \item Analyzing the Environmental Quenching Probability as a function of stellar mass we obtained a relation in which less massive galaxies are more likely to be quenched by the current environment. From our sample of \sami dwarfs only 36$\pm$9\% of them are compatible with being quenched by the environment, while for the \atlas giant galaxies, the percentage falls down to 3$\pm$3\%.
  \item The simulated galaxies of \tng show a different behaviour, with the quenching probability decreasing with decreasing stellar mass. This, in fact, corresponds to massive simulated galaxies being younger than the observational ones, while less massive simulated galaxies are older than our observed sample.
  \item The observed disparity between the trends of less massive galaxies in the simulated and observed samples highlights the intricate interplay of various physical processes governing galaxy evolution. Though we have gained valuable insights into the quenching of dwarf galaxies, further investigation is essential to discern the specific mechanisms responsible for the distinct behaviours observed in both our simulations and real-world data. Such investigation could dive deeper into the role of feedback mechanisms and the influence of the surrounding environment on quenching processes, contributing to a more comprehensive understanding of galaxy evolution.
  \item  Our comparison between observations and simulations underscores the importance of considering sample selections, volume, quenching definition, and projection effects. Caution must be exercised when relying on phase-space parametrization results, especially with smaller sample sizes that may not yield statistical conclusions.  However, this was tested and did not appear to be a major issue. Future studies could address these limitations by exploring larger sample sizes and employing alternative analysis methods. Such endeavours will provide a more robust foundation for interpreting and contextualizing the implications of our findings.
\end{enumerate}

In conclusion, our analysis of observed galaxies combined with simulations reinforces the knowledge of trends between local quenching probability and stellar mass and also shows important inconsistency between observation and simulations. While the observed galaxies exhibit a gradual increase in quenching probability for less massive galaxies, the simulated galaxies show a contrasting decline. This disparity underscores the need for comprehensive theoretical frameworks and a deeper understanding of the intricate processes that shape the evolution of galaxies, particularly for the dwarf population.

\section*{Data availability}
The reduced data underlying this article will be available through \citetalias{Romero-Gomez_2023} in CDS. The raw data is publicly available in the AAT data archive.

\begin{acknowledgements}

JRG and JALA are supported by the Spanish Ministry of Education, Culture and Sports under grant AYA2017-83204-P and by the Spanish Ministerio de Ciencia e Innovaci\'on y Universidades by the grant PID2020-119342GB-I00.

We would like to thank {\bf{Ignacio Martín Navarro and Katja Fahrion}} for kindly giving us their SFHs of the Fornax3D galaxies.

For the analysis, we have used Python \href{http://www.python.org}{http://www.python.org}; Matplotlib \citep{Hunter2007}, a suite of open source python modules that provide a framework for creating scientific plots; and Astropy, a community-developed core Python package for Astronomy \citep{Astropy2013}.
\end{acknowledgements}



\bibliographystyle{aa}
\bibliography{References} 

\begin{thebibliography}{106}
\expandafter\ifx\csname natexlab\endcsname\relax\def\natexlab#1{#1}\fi

\bibitem[{{Aguerri} {et~al.}(2005){Aguerri}, {Gerhard}, {Arnaboldi}, {Napolitano}, {Castro-Rodriguez}, \& {Freeman}}]{Aguerri2005}
{Aguerri}, J.~A.~L., {Gerhard}, O.~E., {Arnaboldi}, M., {et~al.} 2005, \aj, 129, 2585

\bibitem[{{Aguerri} \& {Gonz{\'a}lez-Garc{\'\i}a}(2009)}]{Aguerri-Garcia2009}
{Aguerri}, J.~A.~L. \& {Gonz{\'a}lez-Garc{\'\i}a}, A.~C. 2009, \aap, 494, 891

\bibitem[{{Angthopo} {et~al.}(2021){Angthopo}, {Negri}, {Ferreras}, {de la Rosa}, {Dalla Vecchia}, \& {Pillepich}}]{overquenching_problem}
{Angthopo}, J., {Negri}, A., {Ferreras}, I., {et~al.} 2021, \mnras, 502, 3685

\bibitem[{{Astropy Collaboration} {et~al.}(2013){Astropy Collaboration}, {Robitaille}, {Tollerud}, {Greenfield}, {Droettboom}, {Bray}, {Aldcroft}, {Davis}, {Ginsburg}, {Price-Whelan}, {Kerzendorf}, {Conley}, {Crighton}, {Barbary}, {Muna}, {Ferguson}, {Grollier}, {Parikh}, {Nair}, {Unther}, {Deil}, {Woillez}, {Conseil}, {Kramer}, {Turner}, {Singer}, {Fox}, {Weaver}, {Zabalza}, {Edwards}, {Azalee Bostroem}, {Burke}, {Casey}, {Crawford}, {Dencheva}, {Ely}, {Jenness}, {Labrie}, {Lim}, {Pierfederici}, {Pontzen}, {Ptak}, {Refsdal}, {Servillat}, \& {Streicher}}]{Astropy2013}
{Astropy Collaboration}, {Robitaille}, T.~P., {Tollerud}, E.~J., {et~al.} 2013, \aap, 558, A33

\bibitem[{{Baldry} {et~al.}(2004{\natexlab{a}}){Baldry}, {Balogh}, {Bower}, {Glazebrook}, \& {Nichol}}]{Baldry2004}
{Baldry}, I.~K., {Balogh}, M.~L., {Bower}, R., {Glazebrook}, K., \& {Nichol}, R.~C. 2004{\natexlab{a}}, in American Institute of Physics Conference Series, Vol. 743, The New Cosmology: Conference on Strings and Cosmology, ed. R.~E. {Allen}, D.~V. {Nanopoulos}, \& C.~N. {Pope}, 106--119

\bibitem[{{Baldry} {et~al.}(2006){Baldry}, {Balogh}, {Bower}, {Glazebrook}, {Nichol}, {Bamford}, \& {Budavari}}]{Baldry2006}
{Baldry}, I.~K., {Balogh}, M.~L., {Bower}, R.~G., {et~al.} 2006, \mnras, 373, 469

\bibitem[{{Baldry} {et~al.}(2004{\natexlab{b}}){Baldry}, {Glazebrook}, {Brinkmann}, {Ivezi{\'c}}, {Lupton}, {Nichol}, \& {Szalay}}]{Baldry2004_cmd}
{Baldry}, I.~K., {Glazebrook}, K., {Brinkmann}, J., {et~al.} 2004{\natexlab{b}}, \apj, 600, 681

\bibitem[{{Bidaran} {et~al.}(2022){Bidaran}, {La Barbera}, {Pasquali}, {Peletier}, {van de Ven}, {Grebel}, {Falc{\'o}n-Barroso}, {Sybilska}, {Gadotti}, \& {Coccato}}]{Bidaran2022}
{Bidaran}, B., {La Barbera}, F., {Pasquali}, A., {et~al.} 2022, arXiv e-prints, arXiv:2207.06977

\bibitem[{{Bidaran} {et~al.}(2023){Bidaran}, {La Barbera}, {Pasquali}, {van de Ven}, {Peletier}, {Falc{\'o}n-Barroso}, {Gadotti}, {Sybilska}, \& {Grebel}}]{Bidaran2023}
{Bidaran}, B., {La Barbera}, F., {Pasquali}, A., {et~al.} 2023, \mnras, 525, 4329

\bibitem[{{Binggeli} {et~al.}(1988){Binggeli}, {Sandage}, \& {Tammann}}]{Binggeli1988}
{Binggeli}, B., {Sandage}, A., \& {Tammann}, G.~A. 1988, \araa, 26, 509

\bibitem[{{Blakeslee} {et~al.}(2009){Blakeslee}, {Jord{\'a}n}, {Mei}, {C{\^o}t{\'e}}, {Ferrarese}, {Infante}, {Peng}, {Tonry}, \& {West}}]{Blakeslee2009}
{Blakeslee}, J.~P., {Jord{\'a}n}, A., {Mei}, S., {et~al.} 2009, \apj, 694, 556

\bibitem[{{B{\"o}hringer} {et~al.}(1994){B{\"o}hringer}, {Briel}, {Schwarz}, {Voges}, {Hartner}, \& {Tr{\"u}mper}}]{Boringuer_VirgoXray1994}
{B{\"o}hringer}, H., {Briel}, U.~G., {Schwarz}, R.~A., {et~al.} 1994, \nat, 368, 828

\bibitem[{{Boselli} {et~al.}(2022){Boselli}, {Fossati}, \& {Sun}}]{Boselli2022}
{Boselli}, A., {Fossati}, M., \& {Sun}, M. 2022, \aapr, 30, 3

\bibitem[{{Boselli} \& {Gavazzi}(2006)}]{Boselli2006}
{Boselli}, A. \& {Gavazzi}, G. 2006, \pasp, 118, 517

\bibitem[{{Boselli} \& {Gavazzi}(2014)}]{Boselli2014_}
{Boselli}, A. \& {Gavazzi}, G. 2014, \aapr, 22, 74

\bibitem[{{Boselli} {et~al.}(2014){Boselli}, {Voyer}, {Boissier}, {Cucciati}, {Consolandi}, {Cortese}, {Fumagalli}, {Gavazzi}, {Heinis}, {Roehlly}, \& {Toloba}}]{Boselli2014_enrionment}
{Boselli}, A., {Voyer}, E., {Boissier}, S., {et~al.} 2014, \aap, 570, A69

\bibitem[{{Cappellari}(2017)}]{Capellari2017}
{Cappellari}, M. 2017, \mnras, 466, 798

\bibitem[{{Cappellari} \& {Emsellem}(2004)}]{Capellari2004}
{Cappellari}, M. \& {Emsellem}, E. 2004, \pasp, 116, 138

\bibitem[{{Cappellari} {et~al.}(2011{\natexlab{a}}){Cappellari}, {Emsellem}, {Krajnovi{\'c}}, {McDermid}, {Scott}, {Verdoes Kleijn}, {Young}, {Alatalo}, {Bacon}, {Blitz}, {Bois}, {Bournaud}, {Bureau}, {Davies}, {Davis}, {de Zeeuw}, {Duc}, {Khochfar}, {Kuntschner}, {Lablanche}, {Morganti}, {Naab}, {Oosterloo}, {Sarzi}, {Serra}, \& {Weijmans}}]{Cappellari2011-atlas3d}
{Cappellari}, M., {Emsellem}, E., {Krajnovi{\'c}}, D., {et~al.} 2011{\natexlab{a}}, \mnras, 413, 813

\bibitem[{{Cappellari} {et~al.}(2011{\natexlab{b}}){Cappellari}, {Emsellem}, {Krajnovi{\'c}}, {McDermid}, {Serra}, {Alatalo}, {Blitz}, {Bois}, {Bournaud}, {Bureau}, {Davies}, {Davis}, {de Zeeuw}, {Khochfar}, {Kuntschner}, {Lablanche}, {Morganti}, {Naab}, {Oosterloo}, {Sarzi}, {Scott}, {Weijmans}, \& {Young}}]{Cappellari2011-density}
{Cappellari}, M., {Emsellem}, E., {Krajnovi{\'c}}, D., {et~al.} 2011{\natexlab{b}}, \mnras, 416, 1680

\bibitem[{{Carnall} {et~al.}(2019){Carnall}, {McLure}, {Dunlop}, {Cullen}, {McLeod}, {Wild}, {Johnson}, {Appleby}, {Dav{\'e}}, {Amorin}, {Bolzonella}, {Castellano}, {Cimatti}, {Cucciati}, {Gargiulo}, {Garilli}, {Marchi}, {Pentericci}, {Pozzetti}, {Schreiber}, {Talia}, \& {Zamorani}}]{Carnall2019}
{Carnall}, A.~C., {McLure}, R.~J., {Dunlop}, J.~S., {et~al.} 2019, \mnras, 490, 417

\bibitem[{{Choi} \& {Yi}(2017)}]{Choi_Yi2017}
{Choi}, H. \& {Yi}, S.~K. 2017, \apj, 837, 68

\bibitem[{{Churazov} {et~al.}(2008){Churazov}, {Forman}, {Vikhlinin}, {Tremaine}, {Gerhard}, \& {Jones}}]{Churazov2008}
{Churazov}, E., {Forman}, W., {Vikhlinin}, A., {et~al.} 2008, \mnras, 388, 1062

\bibitem[{{Collins} \& {Read}(2022)}]{Collins2022}
{Collins}, M. L.~M. \& {Read}, J.~I. 2022, Nature Astronomy, 6, 647

\bibitem[{{Cortese} {et~al.}(2021){Cortese}, {Catinella}, \& {Smith}}]{Cortese2021}
{Cortese}, L., {Catinella}, B., \& {Smith}, R. 2021, \pasa, 38, e035

\bibitem[{{Davies} {et~al.}(2019){Davies}, {Robotham}, {Lagos}, {Driver}, {Stevens}, {Bah{\'e}}, {Alpaslan}, {Bremer}, {Brown}, {Brough}, {Bland-Hawthorn}, {Cortese}, {Elahi}, {Grootes}, {Holwerda}, {Ludlow}, {McGee}, {Owers}, \& {Phillipps}}]{Davies2019}
{Davies}, L.~J.~M., {Robotham}, A.~S.~G., {Lagos}, C. d.~P., {et~al.} 2019, \mnras, 483, 5444

\bibitem[{{De Lucia} {et~al.}(2012){De Lucia}, {Weinmann}, {Poggianti}, {Arag{\'o}n-Salamanca}, \& {Zaritsky}}]{DeLucia2012}
{De Lucia}, G., {Weinmann}, S., {Poggianti}, B.~M., {Arag{\'o}n-Salamanca}, A., \& {Zaritsky}, D. 2012, \mnras, 423, 1277

\bibitem[{{den Brok} {et~al.}(2011){den Brok}, {Peletier}, {Valentijn}, {Balcells}, {Carter}, {Erwin}, {Ferguson}, {Goudfrooij}, {Graham}, {Hammer}, {Lucey}, {Trentham}, {Guzm{\'a}n}, {Hoyos}, {Verdoes Kleijn}, {Jogee}, {Karick}, {Marinova}, {Mouhcine}, \& {Weinzirl}}]{denBrok2011}
{den Brok}, M., {Peletier}, R.~F., {Valentijn}, E.~A., {et~al.} 2011, \mnras, 414, 3052

\bibitem[{{Ding} {et~al.}(2023){Ding}, {Zhu}, {van de Ven}, {Coccato}, {Corsini}, {Costantin}, {Fahrion}, {Falc{\'o}n-Barroso}, {Gadotti}, {Iodice}, {Lyubenova}, {Mart{\'\i}n-Navarro}, {McDermid}, {Pinna}, \& {Sarzi}}]{Ding2023}
{Ding}, Y., {Zhu}, L., {van de Ven}, G., {et~al.} 2023, \aap, 672, A84

\bibitem[{{Donnari} {et~al.}(2021){Donnari}, {Pillepich}, {Joshi}, {Nelson}, {Genel}, {Marinacci}, {Rodriguez-Gomez}, {Pakmor}, {Torrey}, {Vogelsberger}, \& {Hernquist}}]{Donnari2021a}
{Donnari}, M., {Pillepich}, A., {Joshi}, G.~D., {et~al.} 2021, \mnras, 500, 4004

\bibitem[{{Drinkwater} {et~al.}(2001){Drinkwater}, {Gregg}, \& {Colless}}]{Drinkwater2001}
{Drinkwater}, M.~J., {Gregg}, M.~D., \& {Colless}, M. 2001, \apjl, 548, L139

\bibitem[{{Eftekhari} {et~al.}(2021){Eftekhari}, {Peletier}, {Scott}, {Bryant}, {Bland-Hawthorn}, {Capaccioli}, {Croom}, {Drinkwater}, {Falc{\'o}n-Barroso}, {Hilker}, {Iodice}, {Lorente}, {Mieske}, {Spavone}, {van de Ven}, \& {Venhola}}]{Eftekhari_2021_fornaxII}
{Eftekhari}, F.~S., {Peletier}, R.~F., {Scott}, N., {et~al.} 2021, \mnras, 497, 1571

\bibitem[{{Fahrion} {et~al.}(2021){Fahrion}, {Lyubenova}, {van de Ven}, {Hilker}, {Leaman}, {Falc{\'o}n-Barroso}, {Bittner}, {Coccato}, {Corsini}, {Gadotti}, {Iodice}, {McDermid}, {Mart{\'\i}n-Navarro}, {Pinna}, {Poci}, {Sarzi}, {de Zeeuw}, \& {Zhu}}]{Fahrion2021}
{Fahrion}, K., {Lyubenova}, M., {van de Ven}, G., {et~al.} 2021, \aap, 650, A137

\bibitem[{{Falc{\'o}n-Barroso} {et~al.}(2011){Falc{\'o}n-Barroso}, {S{\'a}nchez-Bl{\'a}zquez}, {Vazdekis}, {Ricciardelli}, {Cardiel}, {Cenarro}, {Gorgas}, \& {Peletier}}]{Falcon-Barroso2011}
{Falc{\'o}n-Barroso}, J., {S{\'a}nchez-Bl{\'a}zquez}, P., {Vazdekis}, A., {et~al.} 2011, \aap, 532, A95

\bibitem[{{Ferr{\'e}-Mateu} {et~al.}(2018){Ferr{\'e}-Mateu}, {Alabi}, {Forbes}, {Romanowsky}, {Brodie}, {Pandya}, {Mart{\'\i}n-Navarro}, {Bellstedt}, {Wasserman}, {Stone}, \& {Okabe}}]{Ferre2018}
{Ferr{\'e}-Mateu}, A., {Alabi}, A., {Forbes}, D.~A., {et~al.} 2018, \mnras, 479, 4891

\bibitem[{{Fillingham} {et~al.}(2015){Fillingham}, {Cooper}, {Wheeler}, {Garrison-Kimmel}, {Boylan-Kolchin}, \& {Bullock}}]{Fillingham2015_108_mass_therhold}
{Fillingham}, S.~P., {Cooper}, M.~C., {Wheeler}, C., {et~al.} 2015, \mnras, 454, 2039

\bibitem[{{Gavazzi} \& {Scodeggio}(1996)}]{Gavazzi1996}
{Gavazzi}, G. \& {Scodeggio}, M. 1996, \aap, 312, L29

\bibitem[{{Gunn} \& {Gott}(1972)}]{Gunn1972-rampressure}
{Gunn}, J.~E. \& {Gott}, J.~Richard, I. 1972, \apj, 176, 1

\bibitem[{{Haggar} {et~al.}(2020){Haggar}, {Gray}, {Pearce}, {Knebe}, {Cui}, {Mostoghiu}, \& {Yepes}}]{Haggar2020}
{Haggar}, R., {Gray}, M.~E., {Pearce}, F.~R., {et~al.} 2020, \mnras, 492, 6074

\bibitem[{{Heavens} {et~al.}(2004){Heavens}, {Panter}, {Jimenez}, \& {Dunlop}}]{Heavens2004}
{Heavens}, A., {Panter}, B., {Jimenez}, R., \& {Dunlop}, J. 2004, \nat, 428, 625

\bibitem[{{Hirschmann} {et~al.}(2014){Hirschmann}, {De Lucia}, {Wilman}, {Weinmann}, {Iovino}, {Cucciati}, {Zibetti}, \& {Villalobos}}]{Hirschmann2014}
{Hirschmann}, M., {De Lucia}, G., {Wilman}, D., {et~al.} 2014, \mnras, 444, 2938

\bibitem[{{Hunter}(2007)}]{Hunter2007}
{Hunter}, J.~D. 2007, Computing in Science and Engineering, 9, 90

\bibitem[{{Iodice} {et~al.}(2019){Iodice}, {Sarzi}, {Bittner}, {Coccato}, {Costantin}, {Corsini}, {van de Ven}, {de Zeeuw}, {Falc{\'o}n-Barroso}, {Gadotti}, {Lyubenova}, {Mart{\'\i}n-Navarro}, {McDermid}, {Nedelchev}, {Pinna}, {Pizzella}, {Spavone}, \& {Viaene}}]{Iodice2019_fornax3d}
{Iodice}, E., {Sarzi}, M., {Bittner}, A., {et~al.} 2019, \aap, 627, A136

\bibitem[{{Joshi} {et~al.}(2021){Joshi}, {Pillepich}, {Nelson}, {Zinger}, {Marinacci}, {Springel}, {Vogelsberger}, \& {Hernquist}}]{Joshi2021}
{Joshi}, G.~D., {Pillepich}, A., {Nelson}, D., {et~al.} 2021, \mnras, 508, 1652

\bibitem[{{Karachentsev} {et~al.}(2014){Karachentsev}, {Tully}, {Wu}, {Shaya}, \& {Dolphin}}]{Karachentsev2014}
{Karachentsev}, I.~D., {Tully}, R.~B., {Wu}, P.-F., {Shaya}, E.~J., \& {Dolphin}, A.~E. 2014, \apj, 782, 4

\bibitem[{{Kim} {et~al.}(2014){Kim}, {Rey}, {Jerjen}, {Lisker}, {Sung}, {Lee}, {Chung}, {Pak}, {Yi}, \& {Lee}}]{Kim2014}
{Kim}, S., {Rey}, S.-C., {Jerjen}, H., {et~al.} 2014, \apjs, 215, 22

\bibitem[{{Kleiner} {et~al.}(2023){Kleiner}, {Serra}, {Maccagni}, {Raj}, {de Blok}, {J{\'o}zsa}, {Kamphuis}, {Kraan-Korteweg}, {Loi}, {Loni}, {Loubser}, {Moln{\'a}r}, {Oosterloo}, {Peletier}, \& {Pisano}}]{Kleiner2023}
{Kleiner}, D., {Serra}, P., {Maccagni}, F.~M., {et~al.} 2023, \aap, 675, A108

\bibitem[{{Koleva} {et~al.}(2011){Koleva}, {Prugniel}, {De Rijcke}, \& {Zeilinger}}]{Koleva2011}
{Koleva}, M., {Prugniel}, P., {De Rijcke}, S., \& {Zeilinger}, W.~W. 2011, \mnras, 417, 1643

\bibitem[{{Larson} {et~al.}(1980){Larson}, {Tinsley}, \& {Caldwell}}]{Larson1980}
{Larson}, R.~B., {Tinsley}, B.~M., \& {Caldwell}, C.~N. 1980, \apj, 237, 692

\bibitem[{{Lisker}(2009)}]{Lisker2009}
{Lisker}, T. 2009, Astronomische Nachrichten, 330, 1043

\bibitem[{{Lisker} {et~al.}(2006){Lisker}, {Glatt}, {Westera}, \& {Grebel}}]{Lisker2006_gradients}
{Lisker}, T., {Glatt}, K., {Westera}, P., \& {Grebel}, E.~K. 2006, \aj, 132, 2432

\bibitem[{{Liu} {et~al.}(2019){Liu}, {Peng}, {Jord{\'a}n}, {Blakeslee}, {C{\^o}t{\'e}}, {Ferrarese}, \& {Puzia}}]{Liu2019}
{Liu}, Y., {Peng}, E.~W., {Jord{\'a}n}, A., {et~al.} 2019, \apj, 875, 156

\bibitem[{{Maddox} {et~al.}(2019){Maddox}, {Serra}, {Venhola}, {Peletier}, {Loubser}, \& {Iodice}}]{Maddox2019}
{Maddox}, N., {Serra}, P., {Venhola}, A., {et~al.} 2019, \mnras, 490, 1666

\bibitem[{{Marinacci} {et~al.}(2018){Marinacci}, {Vogelsberger}, {Pakmor}, {Torrey}, {Springel}, {Hernquist}, {Nelson}, {Weinberger}, {Pillepich}, {Naiman}, \& {Genel}}]{Marinacci2018}
{Marinacci}, F., {Vogelsberger}, M., {Pakmor}, R., {et~al.} 2018, \mnras, 480, 5113

\bibitem[{{Mart{\'\i}nez-Garc{\'\i}a} {et~al.}(2023){Mart{\'\i}nez-Garc{\'\i}a}, {del Pino}, {{\L}okas}, {van der Marel}, \& {Aparicio}}]{Alberto2023}
{Mart{\'\i}nez-Garc{\'\i}a}, A.~M., {del Pino}, A., {{\L}okas}, E.~L., {van der Marel}, R.~P., \& {Aparicio}, A. 2023, \mnras, 526, 3589

\bibitem[{{Mastropietro} {et~al.}(2005){Mastropietro}, {Moore}, {Mayer}, {Debattista}, {Piffaretti}, \& {Stadel}}]{Mastropietro2005}
{Mastropietro}, C., {Moore}, B., {Mayer}, L., {et~al.} 2005, \mnras, 364, 607

\bibitem[{{McDermid} {et~al.}(2015){McDermid}, {Alatalo}, {Blitz}, {Bournaud}, {Bureau}, {Cappellari}, {Crocker}, {Davies}, {Davis}, {de Zeeuw}, {Duc}, {Emsellem}, {Khochfar}, {Krajnovi{\'c}}, {Kuntschner}, {Morganti}, {Naab}, {Oosterloo}, {Sarzi}, {Scott}, {Serra}, {Weijmans}, \& {Young}}]{McDermid2015}
{McDermid}, R.~M., {Alatalo}, K., {Blitz}, L., {et~al.} 2015, \mnras, 448, 3484

\bibitem[{{Mei} {et~al.}(2007){Mei}, {Blakeslee}, {C{\^o}t{\'e}}, {Tonry}, {West}, {Ferrarese}, {Jord{\'a}n}, {Peng}, {Anthony}, \& {Merritt}}]{Mei2007}
{Mei}, S., {Blakeslee}, J.~P., {C{\^o}t{\'e}}, P., {et~al.} 2007, \apj, 655, 144

\bibitem[{{Michielsen} {et~al.}(2008){Michielsen}, {Boselli}, {Conselice}, {Toloba}, {Whiley}, {Arag{\'o}n-Salamanca}, {Balcells}, {Cardiel}, {Cenarro}, {Gorgas}, {Peletier}, \& {Vazdekis}}]{Michielsen2008}
{Michielsen}, D., {Boselli}, A., {Conselice}, C.~J., {et~al.} 2008, \mnras, 385, 1374

\bibitem[{{Moore} {et~al.}(1998){Moore}, {Lake}, \& {Katz}}]{Moore1998harasment}
{Moore}, B., {Lake}, G., \& {Katz}, N. 1998, \apj, 495, 139

\bibitem[{{Moore} {et~al.}(1999){Moore}, {Lake}, {Quinn}, \& {Stadel}}]{Moore1999}
{Moore}, B., {Lake}, G., {Quinn}, T., \& {Stadel}, J. 1999, \mnras, 304, 465

\bibitem[{{Muzzin} {et~al.}(2008){Muzzin}, {Wilson}, {Lacy}, {Yee}, \& {Stanford}}]{Muzzin2008}
{Muzzin}, A., {Wilson}, G., {Lacy}, M., {Yee}, H.~K.~C., \& {Stanford}, S.~A. 2008, \apj, 686, 966

\bibitem[{{Naiman} {et~al.}(2018){Naiman}, {Pillepich}, {Springel}, {Ramirez-Ruiz}, {Torrey}, {Vogelsberger}, {Pakmor}, {Nelson}, {Marinacci}, {Hernquist}, {Weinberger}, \& {Genel}}]{Naiman2018}
{Naiman}, J.~P., {Pillepich}, A., {Springel}, V., {et~al.} 2018, \mnras, 477, 1206

\bibitem[{{Nelson} {et~al.}(2019){Nelson}, {Pillepich}, {Springel}, {Pakmor}, {Weinberger}, {Genel}, {Torrey}, {Vogelsberger}, {Marinacci}, \& {Hernquist}}]{Nelson2019b}
{Nelson}, D., {Pillepich}, A., {Springel}, V., {et~al.} 2019, \mnras, 490, 3234

\bibitem[{{Nelson} {et~al.}(2018){Nelson}, {Pillepich}, {Springel}, {Weinberger}, {Hernquist}, {Pakmor}, {Genel}, {Torrey}, {Vogelsberger}, {Kauffmann}, {Marinacci}, \& {Naiman}}]{Nelson2018}
{Nelson}, D., {Pillepich}, A., {Springel}, V., {et~al.} 2018, \mnras, 475, 624

\bibitem[{{Panter} {et~al.}(2007){Panter}, {Jimenez}, {Heavens}, \& {Charlot}}]{Panter2007}
{Panter}, B., {Jimenez}, R., {Heavens}, A.~F., \& {Charlot}, S. 2007, \mnras, 378, 1550

\bibitem[{{Parikh} {et~al.}(2024){Parikh}, {Saglia}, {Thomas}, {Mehrgan}, {Bender}, \& {Maraston}}]{Parikh2024}
{Parikh}, T., {Saglia}, R., {Thomas}, J., {et~al.} 2024, \mnras, 528, 7338

\bibitem[{{Pasquali} {et~al.}(2019){Pasquali}, {Smith}, {Gallazzi}, {De Lucia}, {Zibetti}, {Hirschmann}, \& {Yi}}]{Pasquali2019}
{Pasquali}, A., {Smith}, R., {Gallazzi}, A., {et~al.} 2019, \mnras, 484, 1702

\bibitem[{{Peletier}(1989)}]{Reynier1989PhD}
{Peletier}, R.~F. 1989, PhD thesis, -

\bibitem[{{Peng} {et~al.}(2010){Peng}, {Lilly}, {Kova{\v{c}}}, {Bolzonella}, {Pozzetti}, {Renzini}, {Zamorani}, {Ilbert}, {Knobel}, {Iovino}, {Maier}, {Cucciati}, {Tasca}, {Carollo}, {Silverman}, {Kampczyk}, {de Ravel}, {Sanders}, {Scoville}, {Contini}, {Mainieri}, {Scodeggio}, {Kneib}, {Le F{\`e}vre}, {Bardelli}, {Bongiorno}, {Caputi}, {Coppa}, {de la Torre}, {Franzetti}, {Garilli}, {Lamareille}, {Le Borgne}, {Le Brun}, {Mignoli}, {Perez Montero}, {Pello}, {Ricciardelli}, {Tanaka}, {Tresse}, {Vergani}, {Welikala}, {Zucca}, {Oesch}, {Abbas}, {Barnes}, {Bordoloi}, {Bottini}, {Cappi}, {Cassata}, {Cimatti}, {Fumana}, {Hasinger}, {Koekemoer}, {Leauthaud}, {Maccagni}, {Marinoni}, {McCracken}, {Memeo}, {Meneux}, {Nair}, {Porciani}, {Presotto}, \& {Scaramella}}]{Peng2010b}
{Peng}, Y.-j., {Lilly}, S.~J., {Kova{\v{c}}}, K., {et~al.} 2010, \apj, 721, 193

\bibitem[{{Pillepich} {et~al.}(2018{\natexlab{a}}){Pillepich}, {Nelson}, {Hernquist}, {Springel}, {Pakmor}, {Torrey}, {Weinberger}, {Genel}, {Naiman}, {Marinacci}, \& {Vogelsberger}}]{Pillepich2018b}
{Pillepich}, A., {Nelson}, D., {Hernquist}, L., {et~al.} 2018{\natexlab{a}}, \mnras, 475, 648

\bibitem[{{Pillepich} {et~al.}(2019){Pillepich}, {Nelson}, {Springel}, {Pakmor}, {Torrey}, {Weinberger}, {Vogelsberger}, {Marinacci}, {Genel}, {van der Wel}, \& {Hernquist}}]{Pillepich2019}
{Pillepich}, A., {Nelson}, D., {Springel}, V., {et~al.} 2019, \mnras, 490, 3196

\bibitem[{{Pillepich} {et~al.}(2018{\natexlab{b}}){Pillepich}, {Springel}, {Nelson}, {Genel}, {Naiman}, {Pakmor}, {Hernquist}, {Torrey}, {Vogelsberger}, {Weinberger}, \& {Marinacci}}]{Pillepich2018a}
{Pillepich}, A., {Springel}, V., {Nelson}, D., {et~al.} 2018{\natexlab{b}}, \mnras, 473, 4077

\bibitem[{{Pimbblet}(2011)}]{Pimbblet2011_blacksplash}
{Pimbblet}, K.~A. 2011, \mnras, 411, 2637

\bibitem[{{Pintos-Castro} {et~al.}(2019){Pintos-Castro}, {Yee}, {Muzzin}, {Old}, \& {Wilson}}]{Pintos2019}
{Pintos-Castro}, I., {Yee}, H.~K.~C., {Muzzin}, A., {Old}, L., \& {Wilson}, G. 2019, \apj, 876, 40

\bibitem[{{Planck Collaboration} {et~al.}(2016){Planck Collaboration}, {Ade}, {Aghanim}, {Arnaud}, {Ashdown}, {Aumont}, {Baccigalupi}, {Banday}, {Barreiro}, {Bartlett}, {Bartolo}, {Battaner}, {Battye}, {Benabed}, {Beno{\^\i}t}, {Benoit-L{\'e}vy}, {Bernard}, {Bersanelli}, {Bielewicz}, {Bock}, {Bonaldi}, {Bonavera}, {Bond}, {Borrill}, {Bouchet}, {Boulanger}, {Bucher}, {Burigana}, {Butler}, {Calabrese}, {Cardoso}, {Catalano}, {Challinor}, {Chamballu}, {Chary}, {Chiang}, {Chluba}, {Christensen}, {Church}, {Clements}, {Colombi}, {Colombo}, {Combet}, {Coulais}, {Crill}, {Curto}, {Cuttaia}, {Danese}, {Davies}, {Davis}, {de Bernardis}, {de Rosa}, {de Zotti}, {Delabrouille}, {D{\'e}sert}, {Di Valentino}, {Dickinson}, {Diego}, {Dolag}, {Dole}, {Donzelli}, {Dor{\'e}}, {Douspis}, {Ducout}, {Dunkley}, {Dupac}, {Efstathiou}, {Elsner}, {En{\ss}lin}, {Eriksen}, {Farhang}, {Fergusson}, {Finelli}, {Forni}, {Frailis}, {Fraisse}, {Franceschi}, {Frejsel}, {Galeotta}, {Galli}, {Ganga}, {Gauthier}, {Gerbino}, {Ghosh}, {Giard},
  {Giraud-H{\'e}raud}, {Giusarma}, {Gjerl{\o}w}, {Gonz{\'a}lez-Nuevo}, {G{\'o}rski}, {Gratton}, {Gregorio}, {Gruppuso}, {Gudmundsson}, {Hamann}, {Hansen}, {Hanson}, {Harrison}, {Helou}, {Henrot-Versill{\'e}}, {Hern{\'a}ndez-Monteagudo}, {Herranz}, {Hildebrandt}, {Hivon}, {Hobson}, {Holmes}, {Hornstrup}, {Hovest}, {Huang}, {Huffenberger}, {Hurier}, {Jaffe}, {Jaffe}, {Jones}, {Juvela}, {Keih{\"a}nen}, {Keskitalo}, {Kisner}, {Kneissl}, {Knoche}, {Knox}, {Kunz}, {Kurki-Suonio}, {Lagache}, {L{\"a}hteenm{\"a}ki}, {Lamarre}, {Lasenby}, {Lattanzi}, {Lawrence}, {Leahy}, {Leonardi}, {Lesgourgues}, {Levrier}, {Lewis}, {Liguori}, {Lilje}, {Linden-V{\o}rnle}, {L{\'o}pez-Caniego}, {Lubin}, {Mac{\'\i}as-P{\'e}rez}, {Maggio}, {Maino}, {Mandolesi}, {Mangilli}, {Marchini}, {Maris}, {Martin}, {Martinelli}, {Mart{\'\i}nez-Gonz{\'a}lez}, {Masi}, {Matarrese}, {McGehee}, {Meinhold}, {Melchiorri}, {Melin}, {Mendes}, {Mennella}, {Migliaccio}, {Millea}, {Mitra}, {Miville-Desch{\^e}nes}, {Moneti}, {Montier}, {Morgante}, {Mortlock},
  {Moss}, {Munshi}, {Murphy}, {Naselsky}, {Nati}, {Natoli}, {Netterfield}, {N{\o}rgaard-Nielsen}, {Noviello}, {Novikov}, {Novikov}, {Oxborrow}, {Paci}, {Pagano}, {Pajot}, {Paladini}, {Paoletti}, {Partridge}, {Pasian}, {Patanchon}, {Pearson}, {Perdereau}, {Perotto}, {Perrotta}, {Pettorino}, {Piacentini}, {Piat}, {Pierpaoli}, {Pietrobon}, {Plaszczynski}, {Pointecouteau}, {Polenta}, {Popa}, {Pratt}, {Pr{\'e}zeau}, {Prunet}, {Puget}, {Rachen}, {Reach}, {Rebolo}, {Reinecke}, {Remazeilles}, {Renault}, {Renzi}, {Ristorcelli}, {Rocha}, {Rosset}, {Rossetti}, {Roudier}, {Rouill{\'e} d'Orfeuil}, {Rowan-Robinson}, {Rubi{\~n}o-Mart{\'\i}n}, {Rusholme}, {Said}, {Salvatelli}, {Salvati}, {Sandri}, {Santos}, {Savelainen}, {Savini}, {Scott}, {Seiffert}, {Serra}, {Shellard}, {Spencer}, {Spinelli}, {Stolyarov}, {Stompor}, {Sudiwala}, {Sunyaev}, {Sutton}, {Suur-Uski}, {Sygnet}, {Tauber}, {Terenzi}, {Toffolatti}, {Tomasi}, {Tristram}, {Trombetti}, {Tucci}, {Tuovinen}, {T{\"u}rler}, {Umana}, {Valenziano}, {Valiviita}, {Van Tent},
  {Vielva}, {Villa}, {Wade}, {Wandelt}, {Wehus}, {White}, {White}, {Wilkinson}, {Yvon}, {Zacchei}, \& {Zonca}}]{Planck2016}
{Planck Collaboration}, {Ade}, P.~A.~R., {Aghanim}, N., {et~al.} 2016, \aap, 594, A13

\bibitem[{{Quilis} {et~al.}(2000){Quilis}, {Moore}, \& {Bower}}]{Quilis2000}
{Quilis}, V., {Moore}, B., \& {Bower}, R. 2000, Science, 288, 1617

\bibitem[{{Rhee} {et~al.}(2017){Rhee}, {Smith}, {Choi}, {Yi}, {Jaff{\'e}}, {Candlish}, \& {S{\'a}nchez-J{\'a}nssen}}]{Rhee2017_caustic}
{Rhee}, J., {Smith}, R., {Choi}, H., {et~al.} 2017, \apj, 843, 128

\bibitem[{{Romero-G{\'o}mez} {et~al.}(2024){Romero-G{\'o}mez}, {Aguerri}, {Peletier}, {Mieske}, {van de Ven}, \& {Falc{\'o}n-Barroso}}]{Romero-Gomez_2023}
{Romero-G{\'o}mez}, J., {Aguerri}, J.~A.~L., {Peletier}, R.~F., {et~al.} 2024, \mnras, 527, 9715

\bibitem[{{Romero-G{\'o}mez} {et~al.}(2023){Romero-G{\'o}mez}, {Peletier}, {Aguerri}, {Mieske}, {Scott}, {Bland-Hawthorn}, {Bryant}, {Croom}, {Eftekhari}, {Falc{\'o}n-Barroso}, {Hilker}, {van de Ven}, \& {Venhola}}]{Romero-Gomez_2022}
{Romero-G{\'o}mez}, J., {Peletier}, R.~F., {Aguerri}, J.~A.~L., {et~al.} 2023, \mnras, 522, 130

\bibitem[{{Ry{\'s}} {et~al.}(2015){Ry{\'s}}, {Koleva}, {Falc{\'o}n-Barroso}, {Vazdekis}, {Lisker}, {Peletier}, \& {van de Ven}}]{Rys2015}
{Ry{\'s}}, A., {Koleva}, M., {Falc{\'o}n-Barroso}, J., {et~al.} 2015, \mnras, 452, 1888

\bibitem[{{Sales} {et~al.}(2007){Sales}, {Navarro}, {Abadi}, \& {Steinmetz}}]{Sales2007}
{Sales}, L.~V., {Navarro}, J.~F., {Abadi}, M.~G., \& {Steinmetz}, M. 2007, \mnras, 379, 1475

\bibitem[{{Salim} {et~al.}(2007){Salim}, {Rich}, {Charlot}, {Brinchmann}, {Johnson}, {Schiminovich}, {Seibert}, {Mallery}, {Heckman}, {Forster}, {Friedman}, {Martin}, {Morrissey}, {Neff}, {Small}, {Wyder}, {Bianchi}, {Donas}, {Lee}, {Madore}, {Milliard}, {Szalay}, {Welsh}, \& {Yi}}]{Salim2007}
{Salim}, S., {Rich}, R.~M., {Charlot}, S., {et~al.} 2007, \apjs, 173, 267

\bibitem[{{S{\'a}nchez-Bl{\'a}zquez} {et~al.}(2006){S{\'a}nchez-Bl{\'a}zquez}, {Peletier}, {Jim{\'e}nez-Vicente}, {Cardiel}, {Cenarro}, {Falc{\'o}n-Barroso}, {Gorgas}, {Selam}, \& {Vazdekis}}]{Sanchez-Blazquez2006}
{S{\'a}nchez-Bl{\'a}zquez}, P., {Peletier}, R.~F., {Jim{\'e}nez-Vicente}, J., {et~al.} 2006, \mnras, 371, 703

\bibitem[{{Sandage}(1986)}]{Sandage1986}
{Sandage}, A. 1986, \aap, 161, 89

\bibitem[{{Sarzi} {et~al.}(2018){Sarzi}, {Iodice}, {Coccato}, {Corsini}, {de Zeeuw}, {Falc{\'o}n-Barroso}, {Gadotti}, {Lyubenova}, {McDermid}, {van de Ven}, {Fahrion}, {Pizzella}, \& {Zhu}}]{Sarzi2018}
{Sarzi}, M., {Iodice}, E., {Coccato}, L., {et~al.} 2018, \aap, 616, A121

\bibitem[{{Schindler} {et~al.}(1999){Schindler}, {Binggeli}, \& {B{\"o}hringer}}]{Schindler1999}
{Schindler}, S., {Binggeli}, B., \& {B{\"o}hringer}, H. 1999, \aap, 343, 420

\bibitem[{{Scott} {et~al.}(2020){Scott}, {Eftekhari}, {Peletier}, {Bryant}, {Bland-Hawthorn}, {Capaccioli}, {Croom}, {Drinkwater}, {Falc{\'o}n-Barroso}, {Hilker}, {Iodice}, {Lorente}, {Mieske}, {Spavone}, {van de Ven}, \& {Venhola}}]{scott_2020_fornaxI}
{Scott}, N., {Eftekhari}, F.~S., {Peletier}, R.~F., {et~al.} 2020, \mnras, 497, 1571

\bibitem[{{Scoville} {et~al.}(2007){Scoville}, {Abraham}, {Aussel}, {Barnes}, {Benson}, {Blain}, {Calzetti}, {Comastri}, {Capak}, {Carilli}, {Carlstrom}, {Carollo}, {Colbert}, {Daddi}, {Ellis}, {Elvis}, {Ewald}, {Fall}, {Franceschini}, {Giavalisco}, {Green}, {Griffiths}, {Guzzo}, {Hasinger}, {Impey}, {Kneib}, {Koda}, {Koekemoer}, {Lefevre}, {Lilly}, {Liu}, {McCracken}, {Massey}, {Mellier}, {Miyazaki}, {Mobasher}, {Mould}, {Norman}, {Refregier}, {Renzini}, {Rhodes}, {Rich}, {Sanders}, {Schiminovich}, {Schinnerer}, {Scodeggio}, {Sheth}, {Shopbell}, {Taniguchi}, {Tyson}, {Urry}, {Van Waerbeke}, {Vettolani}, {White}, \& {Yan}}]{Scoville2007}
{Scoville}, N., {Abraham}, R.~G., {Aussel}, H., {et~al.} 2007, \apjs, 172, 38

\bibitem[{{Serra} {et~al.}(2023){Serra}, {Maccagni}, {Kleiner}, {Moln{\'a}r}, {Ramatsoku}, {Loni}, {Loi}, {de Blok}, {Bryan}, {Dettmar}, {Frank}, {van Gorkom}, {Govoni}, {Iodice}, {J{\'o}zsa}, {Kamphuis}, {Kraan-Korteweg}, {Loubser}, {Murgia}, {Oosterloo}, {Peletier}, {Pisano}, {Smith}, {Trager}, \& {Verheijen}}]{Serra2023}
{Serra}, P., {Maccagni}, F.~M., {Kleiner}, D., {et~al.} 2023, \aap, 673, A146

\bibitem[{{Smith} {et~al.}(2019){Smith}, {Pacifici}, {Pasquali}, \& {Calder{\'o}n-Castillo}}]{Smith2019}
{Smith}, R., {Pacifici}, C., {Pasquali}, A., \& {Calder{\'o}n-Castillo}, P. 2019, \apj, 876, 145

\bibitem[{{Speagle} {et~al.}(2014){Speagle}, {Steinhardt}, {Capak}, \& {Silverman}}]{Speagle2014}
{Speagle}, J.~S., {Steinhardt}, C.~L., {Capak}, P.~L., \& {Silverman}, J.~D. 2014, \apjs, 214, 15

\bibitem[{{Springel} {et~al.}(2018){Springel}, {Pakmor}, {Pillepich}, {Weinberger}, {Nelson}, {Hernquist}, {Vogelsberger}, {Genel}, {Torrey}, {Marinacci}, \& {Naiman}}]{Springel2018}
{Springel}, V., {Pakmor}, R., {Pillepich}, A., {et~al.} 2018, \mnras, 475, 676

\bibitem[{{Strateva} {et~al.}(2001){Strateva}, {Ivezi{\'c}}, {Knapp}, {Narayanan}, {Strauss}, {Gunn}, {Lupton}, {Schlegel}, {Bahcall}, {Brinkmann}, {Brunner}, {Budav{\'a}ri}, {Csabai}, {Castander}, {Doi}, {Fukugita}, {Gy{\H{o}}ry}, {Hamabe}, {Hennessy}, {Ichikawa}, {Kunszt}, {Lamb}, {McKay}, {Okamura}, {Racusin}, {Sekiguchi}, {Schneider}, {Shimasaku}, \& {York}}]{Strateva2001}
{Strateva}, I., {Ivezi{\'c}}, {\v{Z}}., {Knapp}, G.~R., {et~al.} 2001, \aj, 122, 1861

\bibitem[{{Thomas} {et~al.}(2010){Thomas}, {Maraston}, {Schawinski}, {Sarzi}, \& {Silk}}]{Thomas2010}
{Thomas}, D., {Maraston}, C., {Schawinski}, K., {Sarzi}, M., \& {Silk}, J. 2010, \mnras, 404, 1775

\bibitem[{{Vazdekis} {et~al.}(2010){Vazdekis}, {S{\'a}nchez-Bl{\'a}zquez}, {Falc{\'o}n-Barroso}, {Cenarro}, {Beasley}, {Cardiel}, {Gorgas}, \& {Peletier}}]{Vazdekis2010}
{Vazdekis}, A., {S{\'a}nchez-Bl{\'a}zquez}, P., {Falc{\'o}n-Barroso}, J., {et~al.} 2010, \mnras, 404, 1639

\bibitem[{{Venhola} {et~al.}(2021){Venhola}, {Peletier}, {Salo}, {Laurikainen}, {Janz}, {Haigh}, {Wilkinson}, {Iodice}, {Hilker}, {Mieske}, {Cantiello}, \& {Spavone}}]{Venhola2021}
{Venhola}, A., {Peletier}, R.~F., {Salo}, H., {et~al.} 2021, arXiv e-prints, arXiv:2111.01855

\bibitem[{{Vulcani} {et~al.}(2012){Vulcani}, {Poggianti}, {Fasano}, {Desai}, {Dressler}, {Oemler}, {Calvi}, {D'Onofrio}, \& {Moretti}}]{Vulcani2012}
{Vulcani}, B., {Poggianti}, B.~M., {Fasano}, G., {et~al.} 2012, \mnras, 420, 1481

\bibitem[{{Weinberger} {et~al.}(2017){Weinberger}, {Springel}, {Hernquist}, {Pillepich}, {Marinacci}, {Pakmor}, {Nelson}, {Genel}, {Vogelsberger}, {Naiman}, \& {Torrey}}]{Weinberger2017}
{Weinberger}, R., {Springel}, V., {Hernquist}, L., {et~al.} 2017, \mnras, 465, 3291

\bibitem[{{Weinmann} {et~al.}(2006){Weinmann}, {van den Bosch}, {Yang}, \& {Mo}}]{Weinmann2006}
{Weinmann}, S.~M., {van den Bosch}, F.~C., {Yang}, X., \& {Mo}, H.~J. 2006, \mnras, 366, 2

\bibitem[{{Weisz} {et~al.}(2014){Weisz}, {Dolphin}, {Skillman}, {Holtzman}, {Gilbert}, {Dalcanton}, \& {Williams}}]{Weisz2014}
{Weisz}, D.~R., {Dolphin}, A.~E., {Skillman}, E.~D., {et~al.} 2014, \apj, 789, 147

\bibitem[{{Wetzel} {et~al.}(2013){Wetzel}, {Tinker}, {Conroy}, \& {van den Bosch}}]{Wetzel2013}
{Wetzel}, A.~R., {Tinker}, J.~L., {Conroy}, C., \& {van den Bosch}, F.~C. 2013, \mnras, 432, 336

\bibitem[{{Worthey} {et~al.}(1992){Worthey}, {Faber}, \& {Gonzalez}}]{Worthey1992}
{Worthey}, G., {Faber}, S.~M., \& {Gonzalez}, J.~J. 1992, \apj, 398, 69

\bibitem[{{Xie} {et~al.}(2020){Xie}, {De Lucia}, {Hirschmann}, \& {Fontanot}}]{Xie2020}
{Xie}, L., {De Lucia}, G., {Hirschmann}, M., \& {Fontanot}, F. 2020, \mnras, 498, 4327

\bibitem[{{York} {et~al.}(2000){York}, {Adelman}, {Anderson}, {Anderson}, {Annis}, {Bahcall}, {Bakken}, {Barkhouser}, {Bastian}, {Berman}, {Boroski}, {Bracker}, {Briegel}, {Briggs}, {Brinkmann}, {Brunner}, {Burles}, {Carey}, {Carr}, {Castander}, {Chen}, {Colestock}, {Connolly}, {Crocker}, {Csabai}, {Czarapata}, {Davis}, {Doi}, {Dombeck}, {Eisenstein}, {Ellman}, {Elms}, {Evans}, {Fan}, {Federwitz}, {Fiscelli}, {Friedman}, {Frieman}, {Fukugita}, {Gillespie}, {Gunn}, {Gurbani}, {de Haas}, {Haldeman}, {Harris}, {Hayes}, {Heckman}, {Hennessy}, {Hindsley}, {Holm}, {Holmgren}, {Huang}, {Hull}, {Husby}, {Ichikawa}, {Ichikawa}, {Ivezi{\'c}}, {Kent}, {Kim}, {Kinney}, {Klaene}, {Kleinman}, {Kleinman}, {Knapp}, {Korienek}, {Kron}, {Kunszt}, {Lamb}, {Lee}, {Leger}, {Limmongkol}, {Lindenmeyer}, {Long}, {Loomis}, {Loveday}, {Lucinio}, {Lupton}, {MacKinnon}, {Mannery}, {Mantsch}, {Margon}, {McGehee}, {McKay}, {Meiksin}, {Merelli}, {Monet}, {Munn}, {Narayanan}, {Nash}, {Neilsen}, {Neswold}, {Newberg}, {Nichol}, {Nicinski},
  {Nonino}, {Okada}, {Okamura}, {Ostriker}, {Owen}, {Pauls}, {Peoples}, {Peterson}, {Petravick}, {Pier}, {Pope}, {Pordes}, {Prosapio}, {Rechenmacher}, {Quinn}, {Richards}, {Richmond}, {Rivetta}, {Rockosi}, {Ruthmansdorfer}, {Sandford}, {Schlegel}, {Schneider}, {Sekiguchi}, {Sergey}, {Shimasaku}, {Siegmund}, {Smee}, {Smith}, {Snedden}, {Stone}, {Stoughton}, {Strauss}, {Stubbs}, {SubbaRao}, {Szalay}, {Szapudi}, {Szokoly}, {Thakar}, {Tremonti}, {Tucker}, {Uomoto}, {Vanden Berk}, {Vogeley}, {Waddell}, {Wang}, {Watanabe}, {Weinberg}, {Yanny}, {Yasuda}, \& {SDSS Collaboration}}]{York2000}
{York}, D.~G., {Adelman}, J., {Anderson}, John~E., J., {et~al.} 2000, \aj, 120, 1579

\bibitem[{{Zheng} {et~al.}(2019){Zheng}, {Li}, {Mao}, {Wang}, {Liu}, {Mo}, {Yuan}, {Maraston}, {Thomas}, {Yan}, {Bundy}, {Long}, {Parikh}, {Oyarz{\'u}n}, {Bizyaev}, \& {Lacerna}}]{Zheng2019}
{Zheng}, Z., {Li}, C., {Mao}, S., {et~al.} 2019, \apj, 873, 63

\end{thebibliography}







\label{lastpage}
\end{document}